\documentclass[preprint,letterpaper,tightenlines,nofootinbib]{revtex4}

\usepackage{amsmath}
\usepackage[bookmarks=false]{hyperref}

\newcommand{\eq}[1]{Eq.~\eqref{eq:#1}}
\newcommand{\eqs}[2]{Eqs.~\eqref{eq:#1} and \eqref{eq:#2}}
\renewcommand{\sec}[1]{Sec.~\ref{sec:#1}}
\newcommand{\secs}[2]{Secs.~\ref{sec:#1} and \ref{sec:#2}}
\newcommand{\subsec}[1]{Sec.~\ref{subsec:#1}}

\newcommand{\ie}{i.e.}
\newcommand{\eg}{e.g.}

\newcommand{\abs}[1]{\lvert#1\rvert}
\newcommand{\mae}[3]{\langle#1\lvert#2\rvert#3\rangle}
\newcommand{\Mae}[3]{\bigl\langle#1\bigl\lvert#2\bigr\rvert#3\bigr\rangle}
\newcommand{\bra}[1]{\langle#1\rvert}
\newcommand{\ket}[1]{\lvert#1\rangle}
\newcommand{\ord}[1]{\mathcal{O}(#1)}

\renewcommand{\vec}[1]{\mathbf{#1}}
\newcommand{\ti}[1]{ \tilde{#1}}

\newcommand{\dslash}[1]{ #1 \mspace{-9mu} /}

\newcommand{\deltai}[1]{\delta[\w_{#1} - \cE^0_{#1}]}

\newcommand{\w}{\omega}
\newcommand{\W}{\Omega}

\newcommand{\img}{\mathrm{i}}
\newcommand{\df}{\mathrm{d}}
\newcommand{\Df}{\mathcal{D}}

\newcommand{\vj}{\vec{j}}
\newcommand{\vk}{\vec{k}}
\newcommand{\vn}{\vec{n}}
\newcommand{\vp}{\vec{p}}
\newcommand{\vt}{\vec{t}}

\newcommand{\tp}{\tilde{p}}

\newcommand{\cE}{ \mathcal{E} }
\newcommand{\cL}{ \mathcal{L} }

\newcommand{\cN}{\mathcal{N}}
\newcommand{\cO}{ \mathcal{O} }
\newcommand{\cJ}{ \mathcal{J} }

\newcommand{\cQ}{ \mathcal{Q} }

\newcommand{\mcdot}{\!\cdot\!}

\newcommand{\nslash}{ \dslash{n} }
\newcommand{\bnslash}{ \dslash{\bar{n}} }

\newcommand{\nn}{\nonumber}

\newcommand{\ini}{I}
\newcommand{\Ecm}{E_\mathrm{cm}}

\newcommand{\tr}{\mathrm{tr}}

\newcommand{\jet}{\mathrm{jet}}
\newcommand{\soft}{\mathrm{soft}}
\newcommand{\hemi}{\mathrm{hemi}}
\newcommand{\cone}{\mathrm{cone}}

\newcommand{\Yin}{\ti{Y}}

\newcommand{\Cqq}{C_4}
\newcommand{\Hqq}{H_4}

\newcommand{\citeSCET}{Bauer:2000ew, Bauer:2000yr, Bauer:2001ct, Bauer:2001yt}

\begin{document}

%%%%%%%%%%%%%%%%%%%%%%%%%%%%%%%%%%%%%%%%%%%%%%%%%%%%%%%%%%%%%%%%%%%%%%%%%%%%%%%%
% Title page
%%%%%%%%%%%%%%%%%%%%%%%%%%%%%%%%%%%%%%%%%%%%%%%%%%%%%%%%%%%%%%%%%%%%%%%%%%%%%%%%

\title{Factorization for generic jet production}

\author{Christian W.~Bauer}

\author{Andrew Hornig}

\author{Frank J.~Tackmann}

\affiliation{Ernest Orlando Lawrence Berkeley National Laboratory,
University of California, Berkeley, CA 94720
\vspace{2ex}}

\begin{abstract}

Factorization is the central ingredient in any theoretical prediction for collider experiments. We introduce a factorization formalism that can be applied to any desired observable, like event shapes or jet observables, for any number of jets and a wide range of jet algorithms in leptonic or hadronic collisions. This is achieved by using soft-collinear effective theory to prove the formal factorization of a generic fully-differential cross section in terms of a hard coefficient, and generic jet and soft functions. In this formalism, whether a given observable factorizes in the usual sense, depends on whether it is inclusive enough, so the jet functions can be calculated perturbatively. The factorization formula for any such observable immediately follows from our general result, including the precise definition of the jet and soft functions appropriate for the observable in question. As examples of our formalism, we work out several results in two-jet production for both $e^+ e^-$ and $pp$ collisions.
For the latter, we also comment on how our formalism allows one to treat underlying events and beam remnants.

\end{abstract}

\maketitle

%%%%%%%%%%%%%%%%%%%%%%%%%%%%%%%%%%%%%%%%%%%%%%%%%%%%%%%%%%%%%%%%%%%%%%%%%%%%%%%%
% Main body of the paper

%%%%%%%%%%%%%%%%%%%%%%%%%%%%%%%%%%%%%%%%%%%%%%%%%%%%%%%%%%%%%%%%%%%%%%%%%%%%%%%%
\section{Introduction}
%%%%%%%%%%%%%%%%%%%%%%%%%%%%%%%%%%%%%%%%%%%%%%%%%%%%%%%%%%%%%%%%%%%%%%%%%%%%%%%%

Factorization is the main ingredient in any theoretical prediction for collider experiments. Most factorization theorems are easy to understand intuitively. For example, the most basic factorization theorem for the production of lepton pairs in proton-proton collisions has the form
%%%
\begin{equation}
\label{eq:fact_theorem_simple}
\sigma = \sum_{i,j}  \hat \sigma_{ij} \otimes f_{i/P} \otimes f_{j/P}
\,.\end{equation}
%%%
Here, the partonic cross section $\hat \sigma_{ij}$ describes the production of the two leptons from the two initial state partons $i$ and $j$, while the parton distribution functions $f_{i/P}$ and $f_{j/P}$ describe the probability of finding the partons $i$ and $j$ in the proton. The parton distribution functions and the partonic cross section both depend on the momentum fractions of the partons with respect to the hadrons, and the $\otimes$ denotes the convolution in these variables. For more complicated processes, such as jet production, factorization formulas still exist, but are more complicated than for the Drell-Yan process.

While there is usually a simple intuitive picture leading to factorization theorems like \eq{fact_theorem_simple}, many open questions cannot be answered without a much more detailed understanding of the factorization theorem. First, precise field theoretical definitions of the different elements in terms of matrix elements of operators are required to calculate them systematically. Second, each of these elements depends on a renormalization scale $\mu$, and the precise $\mu$ dependence cannot be obtained from the naive arguments given above. The fact that the final hadronic cross section is independent of $\mu$ allows one to derive renormalization group equations, which can be used to sum large logarithmic terms present in the perturbative results.
Finally, it is important to understand under which circumstances the factorization theorems hold and when they break down. Thus, a more detailed understanding of factorization theorems is mandatory for a theoretical understanding of collider signatures.

Understanding factorization has a long history, and started with the seminal work of Collins, Soper and Sterman (see Refs.~\cite{Collins:1981ta, Collins:1981uk, Collins:1987pm, Collins:1989gx} and references therein). It is instructive to remind the reader about the philosophy of these traditional factorization proofs, and to compare it to factorization proofs using effective fields theories, as discussed in this paper. While the Lagrangian of the strong interaction is given in terms of partonic degrees of freedom, any perturbative calculation of partonic scattering amplitudes gives rise to infrared divergent results. These infrared divergences are a manifestation of the well-known fact that at long distances the strongly interacting degrees of freedom are hadrons, not partons, and that the binding of partons into hadrons is a nonperturbative effect. The infrared divergences in the partonic results are regulated, however, if the dimension of spacetime is chosen to be $D=4-2\epsilon$, and manifest themselves as $1/\epsilon$ poles with $\epsilon < 0$. Thus, for $D\neq 4$ one can calculate the scattering cross section of two partons $m$ and $n$, and by the same intuitive picture as before, one expects that the partonic scattering cross section factorizes as
%%%
\begin{equation}
\label{eq:fact_theorem_partonic}
\sigma_{mn} = \sum_{i,j}\hat \sigma_{ij} \otimes f_{i/m} \otimes f_{j/n}
\,.\end{equation}
%%%
In this case, $f_{i/m}$ [$f_{j/n}$] denotes the probability to find the parton $i$ [$j$] in the parton $m$ [$n$]. This probability has a well-defined expression order by order in perturbation theory, and is infrared divergent in the limit $D \to 4$. Under the \emph{assumption} that any infrared-safe (finite as $D \to 4$) result is the same in the hadronic ($D = 4$) and partonic ($D\neq 4$) theories, the factorization of long distance and short distance physics in the hadronic theory can be proven by showing to all orders in perturbation theory that $\hat\sigma_{ij}$ in the partonic theory is indeed finite in the limit $D \to 4$. Traditional factorization proofs therefore use diagrammatic techniques to show that all infrared-divergent terms in $\sigma_{mn}$ are contained in the partonic distributions $f_{i/m}$ and $f_{j/n}$.

Proofs of factorization theorems in soft-collinear effective theory (SCET)~\cite{\citeSCET}, on the other hand, use a different approach. By construction, the correct effective field theory reproduces the long distance dynamics of the underlying theory in a particular kinematic limit. SCET is constructed to reproduce the long distance physics for processes involving highly energetic particles, and the \emph{assumption} is now that SCET is indeed the correct effective field theory of QCD in this particular kinematic limit. What is important is that this assumption can be tested in perturbation theory.

The physics at short distances, in general, is not properly described by the dynamics of the effective field theory itself, but can be determined by demanding that the effective theory reproduces the underlying theory order by order in perturbation theory. This matching calculation can be performed using partonic degrees of freedom, because the effective theory reproduces the long distance physics of the full theory ensuring that all infrared divergences cancel in the matching calculation. Thus, the effective theory gives a result of the form
%%%
\begin{equation}
\sigma = \sum_{i,j} {\rm SD}_{ij} \otimes {\rm LD}_{ij}
\,,\end{equation}
%%%
where ${\rm SD}_{ij}$ describes the short distance physics governing the scattering of two partons $i$ and $j$ into anything, while ${\rm LD}_{ij}$ describes the long distance probability of two protons to give two partons $i$ and $j$. The final step in the proof of the factorization formula is to show that
%%%
\begin{equation}
{\rm LD}_{ij} = f_{i/P}\otimes f_{j/P}
\end{equation}
%%%
where $f_{i/P}$ are now matrix elements of operators defined in SCET. This is accomplished in SCET by exploiting the dynamics of the effective theory, as will be discussed in much more detail later.

The first factorization proof in SCET was for $B \to D \pi$ decays~\cite{Bauer:2001cu}, but it was soon realized that SCET can be used to reproduce factorization for simple QCD processes with back-to-back jets~\cite{Bauer:2002nz, Bauer:2002ie, Bauer:2003di, Manohar:2003vb}. Recently, there has been progress in studying factorization and the resummation of perturbative corrections for some weighted cross sections in $e^+ e^- \to $ hadrons~\cite{Lee:2006nr, Fleming:2007qr, Fleming:2007xt, Bauer:2008dt, Schwartz:2007ib}, Drell-Yan~\cite{Idilbi:2005ky, Idilbi:2005er, Becher:2007ty}, deep inelastic scattering~\cite{Manohar:2003vb, Chay:2005rz, Manohar:2005az, Chen:2006vd, Becher:2006mr}, and Higgs production~\cite{Gao:2005iu, Idilbi:2005er}.

While such fully inclusive observables have proven to be very useful in capturing generic features of hadronic events, they are not well suited to identify specific short distance processes. For this reason, jet observables are usually considered, in which hadrons that are ``close together'' are collected into jets of particles. The idea is that QCD radiation will turn a single parton produced in a short distance process into a jet of hadrons, such that the total momentum of the jet can be used as a measure of the momentum of the original parton. Thus, jet observables can be used to directly test the underlying short distance process that produced the event, as long as the jets are well separated from the beam axis and their dependence on the underlying event is very small.

Of course, the definition of a jet depends on the precise meaning of ``close together'', and there are many algorithms available which group the final state particles into jets~\cite{Sterman:1977wj, Clavelli:1979md, Chandramohan:1980ry, Clavelli:1981yh, Catani:1991hj, Catani:1992zp, Catani:1993hr, Ellis:1993tq, Huth:1990mi, Ellis:2001aa, Salam:2007xv, Ellis:2007ib}. To calculate jet cross sections perturbatively, we need a jet algorithm that is infrared safe, such that a partonic calculation does not lead to infrared divergences in $D=4$ dimensions. Which algorithm is chosen by experimentalists is usually determined by practical considerations, such as speed and algorithmic robustness.

In this paper, we develop a factorization formalism that can be applied to any desired observable, like event shapes or jet observables. In particular, for $N$-jet production in hadronic collisions, we show that the cross section factorizes into a hard function, $\hat\sigma_{ij,k_1\ldots k_N}$, describing the underlying partonic process to produce $N$ partons, convoluted with $N$ jet functions, $J_{k_i}$, a soft function, $S$, and parton distribution functions, $f_{i,j/P}$,
%%%
\begin{equation}
\label{eq:fact_theorem_further}
\sigma
= \sum_{i,j,k_n}\hat\sigma_{ij,k_1\ldots k_N} \otimes J_{k_1} \otimes \dotsb \otimes J_{k_N} \otimes S
\otimes f_{i/P} \otimes f_{j/P}
\,.\end{equation}
%%%
The jet functions $J_{k_i}$ are the final-state analog of the parton distribution functions. They describe how the final partons from the hard interaction evolve into the observed jets, and contain all dependence on the actual jet algorithm. The soft function $S$ is a nonperturbative object, which describes, for example, how color is rearranged to allow the colored partons to form color-singlet jets. The effective theory allows to give precise field-theoretic definitions of all objects in this factorization formula.
Whether the differential cross section in some observable factorizes in the usual sense depends on whether the observable is dominated by factorizable kinematic configurations and whether it is inclusive enough to allow perturbative calculations of the jet functions.

In \sec{generic}, we first define a generic differential cross section written in terms of functional derivatives, and then show in \sec{observables} how observables are constructed from this generic cross section. In \sec{genericfactorization}, we derive a factorization formula for this differential cross section, which relates it to convolutions over generic building blocks. To focus on the overall structure of the result, this derivation will be schematic in the sense that we will ignore the explicit color and spin structure of the underlying interaction. As a first example, in \sec{ee}, we apply our results to the production of two-jet events in $e^+ e^-$ collisions, including all color and spin information. We reproduce the known results for event shape and hemisphere mass distributions, and obtain results for observables based on cone jet algorithms. In \sec{pp}, we apply our formalism to hadronic collisions. We derive the factorization formula for two-jet production in $pp$ collisions, focussing on the simplest subprocess $qq'\to qq'$. Our conclusions and outlook are presented in \sec{conclusions}.

%%%%%%%%%%%%%%%%%%%%%%%%%%%%%%%%%%%%%%%%%%%%%%%%%%%%%%%%%%%%%%%%%%%%%%%%%%%%%%%%
\section{Energy and three-momentum configuration of an event}
\label{sec:generic}
%%%%%%%%%%%%%%%%%%%%%%%%%%%%%%%%%%%%%%%%%%%%%%%%%%%%%%%%%%%%%%%%%%%%%%%%%%%%%%%%

The differential cross section in any observable $O$ is given by
%%%
\begin{equation}
\label{eq:dsigma_dO}
\frac{\df\sigma}{\df O}
= \frac{1}{2p_I^2} \sum_X\, \abs{ \mae{X}{\cQ}{\ini} }^2 \, (2\pi)^4 \delta^4(p_I - p_X)\,\delta[O - f_O(X)]
\,,\end{equation}
%%%
where $\ket{\ini}$ denotes the initial state containing two particles with total momentum $p_I = (\Ecm,\vec{0})$, $\ket{X}$ denotes an arbitrary final state with total momentum $p_X$, and the sum over $X$ includes a sum over states, as well as all final-state phase-space integrations. Finally, $\cQ$ stands for the relevant operator responsible for the underlying hard interaction.

The function $f_O(X)$ computes the value of the observable for a given final state $X$, and in general depends on the four-momenta of all particles in $X$. The four-momentum configuration of $X$ can be described by its energy configuration $\w_X(\W)$ and three-momentum configuration $\vk_X(\W)$. If X has $n$ particles with four-momenta  $p_i = (E_i, \vec{p}_i)$, we have
%%%
\begin{equation} \label{eq:wXkX}
\w_X(\W) = \sum_{i = 1}^n E_i\, \delta(\W - \W_i)
\,,\qquad
\vk_X(\W) = \sum_{i = 1}^n \vec{p}_i\, \delta(\W - \W_i)
\,,\end{equation}
%%%
where $\W_i \equiv \W(\vec{p}_i)$ is the direction of the three-momentum $\vec{p}_i$. More generally, we can think of $\w(\W)$ and $\vk(\W)$ as the distribution of energy and three-momentum over the solid angle $\W$, as measured experimentally.

To integrate over $\w$ and $\vk$, we define a functional integration measure as usual by discretization. We divide $\W$ into bins $\{\W_k\}$, and define the set of discrete variables $\{\w_k\}$ and $\{\vk_k\}$ as the integrals of $\w(\W)$ and $\vk(\W)$ over the bins $\{\W_k\}$,%
\footnote{This is slightly different from the usual definition of $\Df\phi(x)$ for some field $\phi(x)$, where the discrete variables $\phi_k = \phi(x_k)$ are taken as the values of $\phi$ at the points $x_k$. The difference is an irrelevant overall constant. In our case, taking the integrals is the more natural choice and makes the connection to the usual phase-space integration simpler.}
%%%
\begin{equation}
\w_k = \int_{\W_k}\!\df\W\, \w(\W)
\,,\qquad
\vk_k = \int_{\W_k}\!\df\W\, \vk(\W)
\,.\end{equation}
%%%
Then
%%%
\begin{align}
\label{eq:DwDk}
\Df\w(\W) &\equiv
\Df\w(\W)\, \theta[\w(\W)] = \prod_k \df\w_k\,\theta(\w_k)
\,,\nn\\
\Df\vk(\W)
&\equiv \Df\vk(\W)\, \delta[\W(\vk(\W)) - \W]
= \prod_k \frac{\df^3\vk_k}{(2\pi)^3}\,\delta(\W(\vk_k) - \W_k)
= \prod_k \frac{\abs{\vk_k}^2\,\df\abs{\vk_k}}{(2\pi)^3}
\,.\end{align}
%%%
The $\theta$-functional in $\Df\w(\W)$ restricts $\w(\W)$ to be positive, while the $\delta$-functional in $\Df\vk(\W)$ restricts $\vk(\W)$ to point into the direction $\W$.

The integration measure in \eq{DwDk} still includes unphysical configurations. To only allow physical configurations, we have to include a mass-shell condition. Taking a fixed invariant-mass distribution $\mu(\W)$ as boundary condition, we get
%%%
\begin{align}
\int_{\mu(\W)}\!\Df\w(\W)\,\Df\vk(\W)
&\equiv \int\!\Df\w(\W)\,\Df\vk(\W)\,\delta\bigl[\w(\W)^2 - \vk(\W)^2 - \mu(\W)^2\bigr]
\nn\\
&= \int\!\prod_k \df\w_k\, \frac{\abs{\vk_k}^2\,\df\abs{\vk_k}}{(2\pi)^3}\, \delta(\w_k^2 - \abs{\vk_k}^2 - \mu_k^2)\, \theta(\w_k)
\,,\end{align}
%%%
where $\mu_k = \int_{\W_k}\!\df\W\, \mu(\W)$. This fixes the direction of all particles, but could include different final states $X$, as long as they have the same invariant-mass distribution $\mu(\W)$. If we instead restrict the integration to a state $X$, having $n$ particles with masses $m_i$, we recover the standard $n$-body phase space for $X$,
%%%
\begin{align}
\label{eq:w_x=phase_space}
\int_X\!\Df\w(\W)\,\Df\vk(\W)
&\equiv \int\!\prod_{i = 1}^n\df\W_i\int_{X(\W_1,\ldots,\W_n)}\!\Df\w(\W)\,\Df\vk(\W)
\nn\\
&= \int\!\prod_{i = 1}^n\df\W_i
 \int\!\prod_{k=1}^n \df\w_k\, \frac{\abs{\vk_k}^2\,\df\abs{\vk_k}}{(2\pi)^3}\, \delta(\w_k^2 - \abs{\vk_k}^2 - m_k^2)\, \theta(\w_k)
\nn\\
&= \int\prod_{i = 1}^n \frac{\df^4 p_i}{(2\pi)^3} \,\delta(p_i^2 - m_i^2)\,\theta(p_i^0)
\equiv \int\!\df\Phi_X
\,.\end{align}
%%%
On the right-hand side of the first line, we first integrate $\w(\W)$ and $\vk(\W)$ with the boundary condition that there are exactly $n$ particles with masses $m_i$ moving in the directions $\W_i$, denoted as $X(\W_1,\ldots,\W_n)$, which is then integrated over the particle's positions $\W_i$. In the second line, in the discretization only the integrals over those $n$ bins survive that happen to contain a particle. Together with the $\W_i$ integrations, this reduces to the standard $n$-body phase space for $X$. In the following, we will mostly drop the dependence of $\w(\W)$ and $\vk(\W)$ on $\W$, but one should always keep in mind that $\w$ and $\vk$ are \emph{functions} of $\W$. We will still use square brackets to denote functionals $f[\w]$ and $f[\vk]$.

Returning to \eq{dsigma_dO}, we now assume that $f_O(X)$ does not depend on any internal quantum numbers of $X$, but only on the four-momenta of all particles in $X$.\footnote{Note that most information about internal quantum numbers, such as the number of $b$-jets, is obtained from four-momentum information alone.} In this case, $f_O$ can be written as a functional of the energy and three-momentum configurations,
%%%
\begin{equation}
\label{eq:fO_def}
f_O \equiv f_O[\w,\vk]
\qquad\text{with}\qquad
f_O(X) \equiv f_O[\w_X, \vk_X]
\,.\end{equation}
%%%
We now define an energy-momentum flow operator $\cE^\mu \equiv \cE^\mu(\Omega)$, whose eigenvalues are the energy and three-momentum configurations of the state $\ket{X}$ in \eq{wXkX},
%%%
\begin{equation}
\label{eq:Eflow_action}
\cE^0(\W) \ket{X} = \w_X(\W) \ket{X}
\,,\qquad
\boldsymbol\cE(\W) \ket{X} = \vk_X(\W) \ket{X}
\,.\end{equation}
%%%
The energy flow operator $\cE^0(\W)$ has been used previously, for example to study two-jet event shape distributions~\cite{Korchemsky:1997sy, Korchemsky:1998ev, Korchemsky:1999kt, Belitsky:2001ij, Lee:2006nr, Bauer:2008dt} and jet energy-flow correlations~\cite{Tkachov:1994as, Sveshnikov:1995vi, Tkachov:1995kk, Hofman:2008ar}. In terms of the energy-momentum tensor
%%%
\begin{equation}
T^{\mu\nu}
= \sum_{\phi\in\cL} \frac{\partial\cL}{\partial(\partial_\mu\phi)}\, \partial^\nu \phi - g^{\mu\nu}\cL
\,,\end{equation}
%%%
we can write $\cE^\mu(\W)$ as~\cite{Korchemsky:1997sy, Bauer:2008dt}
%%%
\begin{equation}
\label{eq:Eflow_def}
\cE^\mu(\W) = \lim_{R\to \infty} R^2 \int_0^\infty\!\df t \, \vn^i\, T^{\mu i}(t, R\, \vn)
\,.\end{equation}
%%%
Here, $\vec n \equiv \vec n(\W)$ is the unit three-vector pointing in the direction identified by $\W$. Therefore, $\cE^\mu(\W)$ measures the total four-momentum arriving over time at infinity in the direction $\W$. An expression for $\cE^0(\W)$ similar to \eq{Eflow_def} in terms of an integral over $R$ for $t\to\infty$ was derived in Refs.~\cite{Sveshnikov:1995vi, Cherzor:1997ak}. An explicit proof of \eq{Eflow_def} for $\cE^0(\W)$ for scalars and Dirac fermions can be found in Ref.~\cite{Bauer:2008dt}.

Using \eqs{fO_def}{Eflow_action}, we can write \eq{dsigma_dO} as
%%%
\begin{align}
\frac{\df\sigma}{\df O}
&= \frac{1}{2p_I^2} \sum_X \mae{\ini}{\cQ^\dagger}{X} \mae{X}{\cQ}{\ini}\, (2\pi)^4 \delta^4(p_I - p_X)
\,\delta(O - f_O[\w_X,\vk_X])
\nn\\
&= \frac{1}{2p_I^2} \int\!\Df\w\,\Df\vk\int\!\df^4 x\, \mae{\ini}{\cQ^\dagger(x)\, \delta[\w-\cE^0]\, \delta[\vk-\boldsymbol\cE] \cQ(0)}{\ini}\, \delta(O - f_O[\w,\vk])
\nn\\
&\equiv \int\!\Df\w\,\Df\vk\,\frac{\delta\sigma}{\delta\w\,\delta\vk} \,\delta(O - f_O[\w,\vk])
\,.
\end{align}
%%%
In the second line, we shift $\cQ^\dagger$ to position $x$, turning the momentum conservation into an integral over $x$, and rewrite $\w_X$ and $\vk_X$ in terms of $\cE^\mu$. This removes any explicit dependence on the final state $X$, allowing us to perform the sum over all final states states $\sum_X \ket{X}\bra{X} = 1$.

For the rest of this paper, we will assume for simplicity that $X$ only contains massless particles. The extension to the general case is straightforward. In this case, $f_O \equiv f_O[\w]$ only depends on $\w$, and we can integrate over $\vk$ to find
%%%
\begin{equation}
\label{eq:dsigma_dO2}
\frac{\df\sigma}{\df O}
= \int\!\Df\w\,\frac{\delta\sigma}{\delta\w} \,\delta(O - f_O[\w])
\,,\end{equation}
%%%
where the generic differential cross section $\delta \sigma / \delta \w$ is defined as
%%%
\begin{align}
\label{eq:delsigma_delomega_def}
\frac{\delta \sigma}{\delta\w}
&= \frac{1}{2 p_I^2} \int\!\df^4x\, \mae{\ini}{\cQ^\dagger(x)\,\delta[\w-\cE^0]\, \cQ(0)}{\ini}
\nn\\
&= \frac{1}{2 p_I^2} \int\!\frac{\df^4 p}{(2 \pi)^4}\, \mae{\ini}{\cQ^\dagger(0)\,\delta[\w-\cE^0]\, \cQ(p)}{\ini}
\,.\end{align}
%%%
In the second line, we have written the result in terms of $\cQ(p)$, the Fourier transform of $\cQ(x)$. (To simplify the notation we use the same symbol for operators in position and momentum space and simply distinguish them by their arguments.)

Equation~\eqref{eq:dsigma_dO2} can be regarded as the master formula of our formalism, and its ingredients are the subject of the following sections. We first discuss the functional $f_O[\w]$ in the next section and then the factorization of $\delta\sigma/\delta \w$ in \sec{genericfactorization}. Then in \secs{ee}{pp} we show how to combine these two elements to obtain a factorized form of $\df\sigma/\df O$ for specific processes and observables.

%%%%%%%%%%%%%%%%%%%%%%%%%%%%%%%%%%%%%%%%%%%%%%%%%%%%%%%%%%%%%%%%%%%%%%%%%%%%%%%%
\section{Constructing Observables}
\label{sec:observables}
%%%%%%%%%%%%%%%%%%%%%%%%%%%%%%%%%%%%%%%%%%%%%%%%%%%%%%%%%%%%%%%%%%%%%%%%%%%%%%%%

The form of the functional $f_O[\w]$ depends on the observable under consideration, and in this section we will give a few examples of how to construct $f_O[\w]$ for specific observables. To get used to our notation, we start with the simple example of the total four-momentum of the final state. Next, we consider event shapes, which are fully inclusive observables. Finally, we discuss jet observables, which are less inclusive and defined with respect to a specific jet algorithm.

%===============================================================================
\subsection{Total Four-Momentum}
%===============================================================================

The total energy and three-momentum of the state $X$ are
%%%
\begin{equation}
E_X  = \sum_{i=1}^n E_i = \int\!\df\W\, \w_X(\W)
\,,\qquad
\vec{p}_X = \sum_{i=1}^n \vp_i = \int\!\df\W\, \vn(\W)\, \w_X(\W)
\,,\end{equation}
%%%
where we used that for massless particles $\vk(\W) = \vn(\W)\,\w(\W)$. Hence, we define
%%%
\begin{equation}
\label{eq:Pw}
P^\mu[\w]  = \int\!\df\W \, n^\mu(\W) \,\w(\W)
\,,\end{equation}
%%%
where $n^\mu(\W) = (1, \vn(\W))$. From \eqs{dsigma_dO2}{delsigma_delomega_def} we get
%%%
\begin{align}
\frac{\df\sigma}{\df^4 P}
&= \int\!\Df\w\, \frac{\delta\sigma}{\delta\w}\, \delta^4(P - P[\w])
= \frac{1}{2 p_I^2} \int\!\df^4x\, \mae{\ini}{\cQ^\dagger(x)\, \delta^4(P - \hat{P})\, \cQ(0)}{\ini}
\,.\end{align}
%%%
In the second step we performed the integration over $\w$ and used that (for massless fields) $P^\mu[\cE^0]$ yields the momentum operator $\hat{P}^\mu = (\hat{P}^0, \hat{\vec{P}})$,%
\footnote{To see this explicitly, consider the current $j^\nu(x) \equiv T^{\mu \nu}(x)$ (for fixed $\mu$),
%%%
\begin{equation}
\int\!\df\W\, \cE^\mu(\W)
= \lim_{R\to\infty} \int_0^\infty\df t \int_{\partial S(R)}\df S\, \vn \cdot \vj(t, \vec{x})
= \int_0^\infty\df t \int\df\vec{x}\, \boldsymbol\nabla \cdot \vj(t, \vec{x})
= \lim_{t\to\infty} \int\df\vec{x} \, j^0(t, \vec{x})
\,,\end{equation}
%%%
where in the last step we used current conservation $\partial_\mu j^\mu(x) = 0$.}
%%%
\begin{align}
P^\mu[\cE^0] &= \int\!\df\W\, \cE^\mu(\W)
= \lim_{t\to\infty} \int\!\df\vec{x}\, T^{\mu0}(t,\vec{x})
= \hat{P}^\mu
\,.\end{align}
%%%

%===============================================================================
\subsection{Event Shapes}
%===============================================================================

Event shapes are defined with respect to the thrust axis of an event. Given a final state $X$, one first calculates the thrust axis $\vt \equiv \vt(X)$, which is defined as the unit three-vector $\vt$ that maximizes the sum
%%%
\begin{equation}
\label{eq:t_def}
\sum_{i =1}^n \abs{\vt\cdot\vp_i}
\,,\end{equation}
%%%
which runs over all particles in $X$. Given $\vt$, one then calculates the observable of interest. A generic class of event shapes can be written as
%%%
\begin{equation}
\label{eq:fe_Xdef}
f_e(X) = \frac{1}{\Ecm} \sum_{i = 1}^n g_e(\eta_\vt(p_i))\, \abs{\vp_\vt^T(p_i)}
\,,\end{equation}
%%%
where the rapidity $\eta_\vt$ and transverse momentum $\vp_\vt^T$ are measured with respect to $\vt$. For example, for thrust~\cite{Brandt:1964sa, Farhi:1977sg}, jet broadening~\cite{Catani:1992jc}, and the $C$-parameter~\cite{Ellis:1980wv}, the function $g_e(\eta)$ has the form
%%%
\begin{equation}
g_T(\eta) = e^{-\abs{\eta}}
\,,\qquad
g_B(\eta) = 1
\,,\qquad
g_C(\eta) = \frac{3}{\cosh\eta}
\,.\end{equation}
%%%

The thrust axis can be obtained from the energy configuration of the final state, and can therefore be written as a functional $\vt[\w]$. It is defined (for massless particles) as maximizing the integral
%%%
\begin{equation}
\label{eq:t_wdef}
\int\!\df\W\, \abs{\vt \cdot \vn(\W)}\, \w(\W)
\,,\end{equation}
%%%
which for $\w = \w_X$ reduces to \eq{t_def}. Using $\abs{\vp_\vt^T} = E /\cosh\eta_\vt$, we can write \eq{fe_Xdef} in terms of $\w$
%%%
\begin{equation}
\label{eq:fe_wdef}
f_e[\w] = \int\!\df \vt\, \delta(\vt - \vt[\w]) \, f_{e}[\w; \vt]
\,,\end{equation}
%%%
with
%%%
\begin{equation}
\label{eq:fe_wtdef}
f_e[\w; \vt] = \frac{1}{\Ecm} \int\!\df \W\, g_e(\eta_\vt)\,\frac{\w(\Omega)}{\cosh \eta_\vt}
\,,\end{equation}
%%%
where the solid angle $\W$ is decomposed with respect to the thrust axis as $\W = (\eta_\vt, \phi_\vt)$, and $\eta_\vt = \tanh^{-1}(\cos\theta_\vt)$.

%===============================================================================
\subsection{Jet Observables}
%===============================================================================

%~~~~~~~~~~~~~~~~~~~~~~~~~~~~~~~~~~~~~~~~~~~~~~~~~~~~~~~~~~~~~~~~~~~~~~~~~~~~~~~
\subsubsection{General Features of Jet Algorithms}
%~~~~~~~~~~~~~~~~~~~~~~~~~~~~~~~~~~~~~~~~~~~~~~~~~~~~~~~~~~~~~~~~~~~~~~~~~~~~~~~

A jet algorithm $\cJ$ acting on a final state $X$ returns the set of momenta of all particles in the event, grouped together into the different subsets belonging to each jet plus a set of particles not belonging to any jet, which we take to be soft:
%%%
\begin{equation}
\cJ(X)
= \{ \lbrace p^{\mu} \rbrace_1, \dots, \lbrace p^\mu \rbrace_{\cN}, \lbrace p^\mu \rbrace_s \}
\,.\end{equation}
%%%
In terms of the energy configuration $\w$, we can write the action of the jet algorithm as
%%%
\begin{equation}
\cJ[\w] = \{ \w_1^\jet, \dots, \w^\jet_\cN, \w^\soft \}
\,,\end{equation}
%%%
where $\w_i^\jet$ is the part of $\w$ corresponding to jet $i$, and $\w^\soft$ is the remaining soft part of $\w$ not assigned to any jet, such that
%%%
\begin{equation}
\w = \w^\jet_1 + \dotsb + \w^\jet_\cN + \w^\soft
\,.\end{equation}
%%%

We formally split the action of the jet algorithm into two distinct steps. We first define a quantity $\vj$ that contains all global information about $\w$ required to construct the individual jet configurations $\w_i$ from $\w$. For example, $\vj$ contains the total number of jets and the direction of each jet. That is, $\vj$ is analogous to the thrust axis in the case of event shapes. Hence, a jet algorithm $\cJ$ provides a functional $\vj_\cJ[\w]$, which returns the required information $\vj$ for a given $\w$. Second, we define functionals $\cJ_i[\w; \vj]$ that project out the part of $\w$ belonging to the $i$-th jet. That is, for $\vj = \vj_\cJ[\w]$, by definition
%%%
\begin{equation}
\cJ_i[\w; \vj_\cJ[\w]] = \w^\jet_i
\,,\qquad
\cJ_s[\w; \vj_\cJ[\w]] = \w^\soft
\,.\end{equation}
%%%
We stress that $\cJ_i[\w; \vj]$ only encodes the actual projection, which is completely specified by the specifics of the jet algorithm and the information provided by $\vj$. Thus, for a given $\vj$, $\cJ_i$ can be applied to any $\w$. For example, by definition, $\cJ_i$ satisfies the consistency conditions
%%%
\begin{equation}
\cJ_i\bigl[\w^\jet_j; \vj_\cJ[\w]\bigr] =
\begin{cases}
\w^\jet_i &\text{for} \quad i = j\\
0 &\text{for} \quad i\neq j
\,.\end{cases}
\end{equation}
%%%
For simplicity, we will keep the dependence on $\cJ$ implicit from now on and simply write $\vj[\w]$.

%~~~~~~~~~~~~~~~~~~~~~~~~~~~~~~~~~~~~~~~~~~~~~~~~~~~~~~~~~~~~~~~~~~~~~~~~~~~~~~~
\subsubsection{Construction of Jet Observables}
%~~~~~~~~~~~~~~~~~~~~~~~~~~~~~~~~~~~~~~~~~~~~~~~~~~~~~~~~~~~~~~~~~~~~~~~~~~~~~~~

We can now write a generic jet observable as
%%%
\begin{equation}
f_O[\w] = \int\!\df\vj\,\delta(\vj - \vj[\w])\,f_O[\w; \vj]
\,,\end{equation}
where, in general, $f_O[\w; \vj]$ has the form
%%%
\begin{equation}
f_O[\w; \vj] = f_O[\cJ_1[\w; \vj], \ldots, \cJ_{\cN}[\w; \vj]]
\,,\end{equation}
%%%
and $\cN$ is the total number of jets given by $\vj$. Most jet observables only depend on the total four-momentum of each jet. In this case, using \eq{Pw}%
\footnote{For jet algorithms, \eq{Pw} corresponds to the so-called E-scheme, which defines the total jet momentum as the sum of the particle momenta.},
we have
%%%
\begin{equation} \label{eq:fO_jetP}
f_O[\w; \vj]
= \biggl[ \prod_{i=1}^{\cN} \int\!\df^4 P_i \, \delta^4(P_i- P[\cJ_i[\w;\vj]]) \biggr]
 g_O(P_1, \dots,  P_{\cN})
\,.\end{equation}
%%%
The function $g_O(P_1,\ldots,P_\cN)$ computes the observable of interest from the given jet momenta $P_i$. It is analogous to the function $g_e(\eta)$ for event shapes. Some simple examples would be the total number of jets or the invariant mass of two jets,
%%%
\begin{equation}
g_\cN(P_1,\dots,P_\cN) = \cN
\,,\qquad
g_{m_{ij}}(P_1,\dots,P_\cN) = (P_i + P_j)^2
\,.\end{equation}
%%%
Similar to \eq{fO_jetP}, one can easily define observables depending on additional information about the individual jets, for example, one can imagine observables which depend on a weighted integral of the energies of all particles in a jet.

%~~~~~~~~~~~~~~~~~~~~~~~~~~~~~~~~~~~~~~~~~~~~~~~~~~~~~~~~~~~~~~~~~~~~~~~~~~~~~~~
\subsubsection{Examples of Jet Algorithms}
%~~~~~~~~~~~~~~~~~~~~~~~~~~~~~~~~~~~~~~~~~~~~~~~~~~~~~~~~~~~~~~~~~~~~~~~~~~~~~~~

Of course, in practice, how the action of the jet algorithm is separated into $\vj[\w]$ and $\cJ_i[\w; \vj]$ depends on the actual algorithm, and we will briefly discuss a few examples here. Since most jet algorithms do not have a simple analytic expression for generic final states, it will not be possible to obtain $\vj[\w]$ analytically, either. However, this is not a limitation, because we can always define $\vj[\w]$ by acting with the full jet algorithm on $\w$ and only returning the necessary global information. The more relevant, and perhaps nontrivial, task is to figure out the information required in $\vj$, and to define the projections $\cJ_i[\w; \vj]$ accordingly.

The simplest example is probably the hemisphere jet algorithm \cite{Clavelli:1979md, Chandramohan:1980ry, Clavelli:1981yh}, for which the number of jets is always 2, and the only relevant global information is the axis perpendicular to the plane separating the two hemispheres, which is usually taken to be the thrust axis $\vt$. Hence, we can write $\vj$ as
%%%
\begin{equation}
\vj = \{2; \vt\}
\,,\qquad
\vj[\w] = \{2; \vt[\w] \}
\,,\end{equation}
%%%
where by convention we included the number of jets in $\vj$. The corresponding projections $\cJ_1$ and $\cJ_2$ return the two hemispheres defined by the thrust axis,
%%%
\begin{equation}
\label{eq:hemi_J}
\cJ_{1}[\w; \vj] = \w(\W) \, \theta\bigl(0 \le \theta_\vt < \pi/2 \bigr)
\,,\qquad
\cJ_{2}[\w; \vj] = \w(\W) \, \theta\bigl(\pi/2\le \theta_\vt < \pi \bigr)
\,,\end{equation}
%%%
where $\W \equiv (\theta_\vt, \phi_\vt)$ is given with respect to $\vt$. Note that here we have $\cJ_s[\w,\vj] = 0$. Typical observables constructed from these individual jets are their invariant masses, $g_{M_{1,2}^2}(P_1, P_2) = P_{1,2}^2$. Another class of observables is given by the event shapes in \eq{fe_wtdef}, for which $f_O[\w, \vj] \equiv f_e[\w, \vt]$.

A less trivial example is a cone jet algorithm~\cite{Sterman:1977wj, Huth:1990mi, Ellis:2001aa, Salam:2007xv, Ellis:2007ib}. In this case, the necessary global information returned by $\vj[\w]$ is the number of jets $\cN$, and the direction $\vj_i$ of each jet,
%%%
\begin{equation}
\label{eq:j_cone}
\vj = \{\cN; \vj_1,\dots,\vj_\cN \}
\,.\end{equation}
%%%
For instance, in $e^+ e^-$ collisions one can define an $\cN$-jet final state as one admitting a minimum number $\cN$ directions $\vj_i$ such that the total energy outside an opening half angle $R$ about each direction is less than some fraction $\epsilon$. For $\cN=2$, and with the additional constraint $\vj_1 = - \vj_2$, this is equivalent to the original Sterman-Weinberg jet definition~\cite{Sterman:1977wj}. For a given set $\vj$, the projections are then simply
%%%
\begin{equation} \label{eq:cone_J}
\cJ_i[\w; \vj] = \w(\W) \, \theta(R  - \theta_{\vj_i})
\,,\end{equation}
%%%
where $R$ is the cone radius and $\W \equiv (\theta_{\vj_i}, \phi_{\vj_i})$ is given with respect to $\vj_i$ for each $i$.

As with all observables, we require jet algorithms that are infrared safe, which is not the case for many cone jet algorithms. An example of an infrared-safe cone jet algorithm is the seedless infrared-safe (SIS) cone jet algorithm~\cite{Salam:2007xv}. For illustration of our
method, we will use the snowmass cone algorithm~\cite{Huth:1990mi, Ellis:2007ib} as an example,
however our approach can easily be adapted to the
SIS algorithm or any other infrared-safe jet algorithm.

For hadronic collisions, the variables $(\eta, \phi)$, defined with respect to the beam axis, have simple transformations under boosts along the beam direction, and so are favored over $(\theta, \phi)$. In the Snowmass cone algorithm, jets are defined by cones of constant radius $R$ in $(\eta, \phi)$ space. When applied to massless particles, the directions $\vj_i$ are given by the solutions of
\footnote{For some configurations $\w$, this equation can admit multiple sets of solutions $\vj$, containing a different number of jets. This happens when there are overlapping cones, and one has to decide whether to split or merge these.}
%%%
\begin{equation} \label{eq:j_snowmass}
0 = \int\!\df \W \, \frac{\w(\W)}{\cosh\eta}\, [\W -  \W(\vj_i)]\,
\theta\bigl(R  - \sqrt{[\eta - \eta(\vj_i)]^2 + [\phi - \phi(\vj_i)]^2}\bigr)
\,,\end{equation}
%%%
where $\W = (\eta, \phi)$ is now measured with respect to the beam axis, and $\W(\vj_i) = (\eta(\vj_i), \phi(\vj_i))$ are the coordinates of the $i$-th jet direction. Equation~\eqref{eq:j_snowmass} is the analog of \eq{t_wdef} for the thrust axis. The corresponding projections are
 %%%
\begin{equation} \label{eq:snowmass_J}
\cJ_i[\w; \vj] = \w(\W) \, \theta\bigl(R  - \sqrt{[\eta - \eta(\vj_i)]^2 + [\phi - \phi(\vj_i)]^2}\bigr)
\,.\end{equation}
%%%

Finally, $k_T$ jet algorithms \cite{Catani:1991hj, Catani:1992zp, Catani:1993hr, Ellis:1993tq} also fit our general definition of algorithms. Although their definition contains a cut on some distance measure, the precise size and shape of a jet also depends on the details of the energy configuration $\w$. Hence, there is no simple expression for the projections $\cJ_i[\w; \vj]$ for arbitrary $\w$ and $\vj$. In principle, they are well defined (albeit complicated) for a fixed number of particles, in which case integrals over the energy configuration reduce to normal phase space integrals [see~\eq{w_x=phase_space}]. In practical applications, it is easiest to apply the algorithm numerically.

%%%%%%%%%%%%%%%%%%%%%%%%%%%%%%%%%%%%%%%%%%%%%%%%%%%%%%%%%%%%%%%%%%%%%%%%%%%%%%%%
\section{\boldmath Factorization of $\delta \sigma / \delta \w$}
\label{sec:genericfactorization}
%%%%%%%%%%%%%%%%%%%%%%%%%%%%%%%%%%%%%%%%%%%%%%%%%%%%%%%%%%%%%%%%%%%%%%%%%%%%%%%%

In this section, we prove a factorization theorem for the generic differential distribution $\delta \sigma / \delta \w$ defined in \eq{delsigma_delomega_def}. This will allow us to separate the various scales in the problem and write our result in terms of convolutions over simpler functions, each of which captures only the physics at a certain energy scale. The factorization proof uses arguments similar to those used to prove factorization for event shape distributions in Ref.~\cite{Bauer:2008dt}. The central ingredient in addition to the usual factorization of soft and collinear degrees of freedom in SCET will be the use of the energy flow operator $\cE^\mu(\W)$ defined in \eq{Eflow_def}.

When deriving the factorization formula, we will ignore all color and spin structure of the SCET operators, and denote all collinear fields by $\phi$, regardless of whether they correspond to quarks, anti-quarks or gluons. This schematic notation will allow us to focus on the issues directly related to the proof of factorization. Of course, to obtain the full result for the cross section, the color and spin information has to be included, and we illustrate how this is achieved when discussing explicit examples in \secs{ee}{pp}.

%===============================================================================
\subsection{Matching QCD onto SCET in Momentum Space}
%===============================================================================

Usually, the matching of QCD onto SCET is performed in position space by expanding the relevant QCD operator $\cQ(x)$ in terms of SCET operators $\cO(x)$,
%%%
\begin{equation}
\label{eq:matching_usual}
\cQ(x) = \sum_{\{n_i, \tp_i\}}  e^{- \img \sum\tp_i \cdot x}\, C_{\{n_i\}}(\{\tp_i\})\, \cO_{\{n_i,\tp_i\}}(x)
\,.\end{equation}
%%%
Here, $C_{\{n_i\}}\cO_{\{n_i,\tp_i\}}$ stands for a sum over several SCET operators with the same number of collinear directions, each with its own Wilson coefficient. The Wilson coefficients are determined by taking matrix elements of both sides, and expanding the full-theory matrix elements $\langle \cQ(x)\rangle$ in terms of the matrix elements $\langle \cO(x) \rangle$ evaluated in SCET. Note also that at this point the operators include all incoming and outgoing fields, whether they are strongly interacting or not.

The operator $\cO_{\{n_i,\tp_i\}}(x)$ is written in terms of (gauge-invariant) collinear SCET fields $\phi_{n, \tp}(x)$. Each field depends on a large label momentum $\tp^\mu = \tp^- n^\mu/2 + \tp_\perp^\mu$ with $n^2 = 0$ and $\tp_\perp \sim \ord{\lambda\tp^-}$, and has a residual $x$ dependence corresponding to a residual momentum $k \sim \ord{\lambda^2\tp^-}$, so the total momentum of the field is $p = \tp + k$. Thus, one can think of the fields $\phi_{n, \tp}(x)$ as being written in label momentum space and residual position space. With this interpretation, the sum over all labels $n_i$ and $\tp_i$ in \eq{matching_usual} corresponds to taking the remaining label Fourier transform to convert the right-hand side to full position space.

This separation into discrete label and continuous residual components is conceptually convenient when formulating the effective theory, and means that phase space is divided up as
%%%
\begin{equation}
\int\!\df^4 p = \sum_{n,\tp} \int\!\df^4 k
\,.\end{equation}
%%%
The concrete choice of the discrete labels $n$ and $\tp$ is determined by the external momenta. As is well-known, this choice is arbitrary at subleading order in $\lambda$, which can be exploited to derive constraints from reparametrization invariance~\cite{Chay:2002vy, Manohar:2002fd}.

However, in practical applications, especially with more than one collinear direction, the label choice can easily get obscured during the calculation. One example is four-momentum conservation for two back-to-back jets with collinear momenta $p_1 = \tp_1 + k_1$ and $p_2 = \tp_2 + k_2$,
%%%
\begin{equation}
\int\!\df^4 x\, e^{\img(\tp_1 - \tp_2 + k_1 - k_2)\cdot x}
= \delta_{\tp_1, \tp_2} \int\!\df^4 x\, e^{\img(k_1 - k_2)x}
= \delta_{\tp_1, \tp_2 + \delta_k} \int\!\df^4 x\, e^{\img(k_1 - k_2 + \delta_k)x}
\,,\end{equation}
%%%
where $\delta_k \sim \ord{k_{1,2}}$. Both equations are correct and correspond to different choices of the label momenta. Using the first equality, as is often done, may seem somewhat ad hoc, but one simply makes an implicit choice of, say, $\tp_2$ relative to $\tp_1$. Of course, this is only justified if $\tp_2$ was not already chosen somewhere else. Furthermore, at the end of the day, one often has to recombine some leftover label sums and residual integrations, \eg,
%%%
\begin{equation}
\sum_n \int\!\df k_\perp \sim \int\!\df\W
\,,\end{equation}
%%%
corresponding to unconstrained phase space integrations of external particles.

With several collinear directions, keeping track of all label choices and dealing with leftover label sums and residual integrations quickly becomes very cumbersome. To avoid all of these issues, it is convenient to perform the matching entirely in momentum space, so \eq{matching_usual} becomes
%%%
\begin{equation}
\label{eq:matching_momentum}
\cQ(p)
= \sum_{\{n_i, \tp_i\}} \biggl(\prod_i \int\!\frac{\df^4k_i}{(2\pi)^4} \biggr)\,
  (2 \pi)^4 \delta^4\Bigl(p - \sum_i(\tp_i + k_i)\Bigl)\, C_{\{n_i\}}(\{\tp_i\})\, \cO_{\{n_i,\tp_i\}}(\{k_i\})
\,.\end{equation}
%%%
Here, $\cQ(p)$ is the Fourier transform of $\cQ(x)$, and $\cO_{\{n_i,\tp_i\}}(\{k_i\})$ is written in terms of momentum-space SCET fields, $\phi_{n,\tp}(k)$, which are obtained by taking the remaining residual Fourier transform of $\phi_{n, \tp}(x)$,
%%%
\begin{equation}
\phi_{n,\tp}(k) = \int\!\df^4 x \, e^{\img k \cdot x}\, \phi_{n, \tp}(x)
\,.\end{equation}
%%%

We can imagine that the matching is performed at fixed total momentum $p_i = \tp_i + k_i$ of each field in $\cO_{\{n_i, \tp_i\}}$. We then choose the field labels directly during the matching such that $\vn = \vp / \abs{\vp}$ for each field. With this choice, $\tp^- \equiv p^- = p^0 + \abs{\vp}$, $p^+ \equiv k^+ = p^0 - \abs{\vp}$, and $p_\perp = 0$. This allows us to recombine the label sums and residual integrations in \eq{matching_momentum} into $\df^4 p_i$ integrals%
\footnote{We suppress that, strictly speaking, the integral over $p$ should be restricted to only include collinear momenta, which is equivalent to excluding the zero-bin region~\cite{Manohar:2006nz} from the integral.}
%%%
\begin{align}
\label{eq:p_space}
\sum_{n,\tp} \int\!\frac{\df^4 k}{(2\pi)^4}
\equiv \int\!\frac{\df^4 p}{(2 \pi)^4}
= \int\!\frac{\df p^- \df p^+\, \df \W }{(2\pi)^4}\, \frac{(p^- - p^+)^2}{8}
\,,\end{align}
%%%
where $\W \equiv \W(\vp)$ is the solid angle corresponding to the direction of $\vp$. We also keep the dependence on the labels implicit and simply write the fields in the operator in terms of their total momentum $p$,
%%%
\begin{equation}
\phi(p) \equiv \phi_{n,\tp}(k)
\,,\qquad
\cO(\{p_i\}) \equiv \cO_{\{n_i, \tp_i\}}(\{k_i\})
\,.\end{equation}
%%%
Hence, the final form of the matching becomes
%%%
\begin{equation}
\label{eq:matching_final}
\cQ(p)
= \biggl(\prod_i \int\!\frac{\df^4p_i}{(2\pi)^4} \biggr)\,
  (2 \pi)^4\delta^4\Bigl(p - \sum_i p_i\Bigl)\, C(\{p_i\})\, \cO(\{p_i\})
\,,\end{equation}
%%%
where here and in the following it is understood that the Wilson coefficient $C$ only depends on the directions $\vn_i = \vp_i/\abs{\vp_i}$ and large components $p_i^- = p_i^0 + \vec{p}_i$ of the momenta $p_i$.

%===============================================================================
\subsection{Factorization Proof}
%===============================================================================

Starting from the definition of $\delta\sigma/\delta\w$ in \eq{delsigma_delomega_def}, in the first step we match QCD onto SCET. According to \eq{matching_final}, the matching condition takes the form
%%%
\begin{align}
\label{eq:matching_step1}
 \cQ(p)
 &= \biggl(\prod_{i = a,b,1}^{N} \int\!\frac{\df^4 p_i}{(2 \pi)^4}\biggr)
\, C_N(p_a, p_b; p_1, \dots, p_N)
\nn\\ & \quad
\times \cO_I(p_a, p_b)\, \cO_F(p_1, \dots, p_N)\, (2 \pi)^4 \delta^4\biggl(p - p_a - p_b + \sum_{i=1}^N p_i\biggr)
\,.\end{align}
%%%
Here, $p_a$ and $p_b$ are initial state collinear momenta, and the operator $\cO_I$ is responsible for annihilating the initial state. Similarly, $p_1,\ldots,p_N$ are $N$ final state collinear momenta, and the operator $\cO_F$, defined as
%%%
\begin{align}
\label{eq:Of_def}
% \cO_I(p_a, p_b) = \phi(-p_a)\, \phi(-p_b)
% \,,\qquad
\cO_F(p_1,\ldots,p_N) = \prod_{i = 1}^N \phi^\dagger(p_i)
\,,\end{align}
%%%
is responsible for creating the final state. Equation~\eqref{eq:matching_step1} is valid in any region of multi-body phase space that is dominated by $N$ jets of collinear particles, corresponding to $N$ collinear directions, that are well separated from each other and the beam axis, i.e., the initial collinear directions $p_{a,b}$. The dominant power corrections to \eq{matching_step1} scale like $p_i^2/ p_i\cdot p_j$.

The different collinear fields in $\cO_I$ and $\cO_F$ interact with each other only through the exchange of soft gluons. These interactions are eliminated to all orders in $\alpha_s$ and leading order in the power counting by the usual field redefinition in SCET~\cite{Bauer:2001yt},
%%%
\begin{equation}
\label{eq:BPS}
\phi_{n,\tp}(x) = Y_n(x)\, \phi_{n,\tp}^{(0)}(x)
\,,\end{equation}
%%%
where $Y_n(x)$ denotes the appropriate soft Wilson line along the direction $n$ in the color representation of $\phi_{n,\tp}(x)$. For color singlet fields, $Y(x)=1$.
As usual, we will drop the superscript on the redefined fields and operators henceforth. In terms of the redefined fields, the matching condition in \eq{matching_step1} takes the form
%%%
\begin{align}
\label{eq:matching_step2}
\cQ(p)
 &= \biggl(\prod_{i = a,b,1}^{N} \int\!\frac{\df^4 p_i}{(2 \pi)^4}\biggr)
\, C_N(p_a, p_b; p_1, \dots, p_N)\,\int\frac{\df^4 k_s}{(2 \pi)^4}
\nn\\ & \quad
\times
\cO_I(p_a, p_b)\, \cO_F(p_1, \dots, p_N)\, \cO_S(k_s)\,
(2 \pi)^4 \delta^4\biggl(p - p_a - p_b + k_s + \sum_{i=1}^N p_i\biggr)
\,,\end{align}
%%%
where the soft operator $\cO_S(k_s)$ contains the Fourier transform of the time-ordered product of all soft Wilson lines,
%%%
\begin{align}
\cO_S(k_s)
&= \int\!\df^4 x\,e^{-\img k_s\cdot x}\, T\biggl[Y_{n_a}(x)\, Y_{n_b}(x)\biggl(\prod_{i=1}^N Y_{n_i}^\dagger(x)\biggr)\biggr]
\,.\end{align}
%%%

Having factored the operator $\cQ(p)$, we next move our attention to the $\delta[\w - \cE^0]$ term in~\eq{delsigma_delomega_def}. After the field redefinition, the leading-order SCET Lagrangian with $N+2$ collinear directions can be written as
%%%
\begin{equation} \label{eq:LSCET}
\cL_\mathrm{SCET} = \sum_{i = a,b,1}^N \cL_i + \cL_s
\,,\end{equation}
%%%
where $\cL_i$ only contains collinear fields in the direction $n_i$, and $\cL_s$ is the purely soft Lagrangian. Since the energy-momentum flow operator, defined in \eq{Eflow_def}, is linear in the Lagrangian of the theory, we have
%%%
\begin{equation} \label{eq:Eflowlinearity}
\cE^\mu(\W) = \sum_{i=a,b,1}^N \cE^\mu_i(\W) + \cE^\mu_s(\W)
\,,\end{equation}
%%%
where $\cE^\mu_{i,s}(\W)$ is defined analogously to \eq{Eflow_def}, but  using the energy-momentum tensor obtained from the Lagrangian $\cL_{i,s}$ only. Thus, $\cE_i^\mu(\W)$ describes the energy-momentum flow in the $i$-th collinear sector, while $\cE^\mu_s(\W)$ describes the remaining soft energy-momentum flow. This allows us to write
%%%
\begin{equation} \label{eq:Eflowfactorization}
\delta[\w - \cE^0] =
\biggl(\prod_{i=a,b,1}^N \int\!\Df\w_i \, \delta[\w_i - \cE^0_i]\biggr)
\int\!\Df\w_s\, \delta[\w_s - \cE^0_s]\,\delta\biggl[\w - \w_s - \sum_{i=a,b,1}^N \w_i\biggr]
\,.\end{equation}
%%%

Combining \eqs{Eflowfactorization}{matching_step2} with \eq{delsigma_delomega_def}, and letting
$\Phi_N = \{p_a, p_b; p_1,\ldots,p_N\}$ denote a point in $(2\to N)$-body phase space (with $\df\Phi_N$ the corresponding phase space measure), $\delta\sigma/\delta\omega$ can be written as
%%%
\begin{align}
\label{eq:matching_step3}
\frac{\delta\sigma}{\delta\w} &=
\frac{1}{2 p_I^2}
\int\!\df\Phi'_N\,\df\Phi_N\, C_N^*(\Phi'_N)\,C_N(\Phi_N)\,
\int\frac{\df^4 k_s'}{(2 \pi)^4}\,\frac{\df^4 k_s}{(2 \pi)^4}\,
\biggl(\prod_{i=a,b,1}^N \int\!\Df\w_i \biggr)\int\!\Df\w_s
\nn\\ & \quad \times
\Big \langle I \Big | (\cO_I \cO_F \cO_S)^\dagger(\Phi_N', k_s')\,
\biggl(\prod_{i=a,b,1}^N \delta[\w_i - \cE^0_i]\biggr) \delta[\w_s - \cE^0_s]\,
(\cO_I \cO_F \cO_S)(\Phi_N, k_s)\Big | I \Big \rangle
\nn\\ & \quad \times
(2 \pi)^4 \delta^4(\Phi_N' - k_s)\,\delta\biggl[\w - \w_s - \sum_{i=a,b,1}^N \w_i\biggr]
\,.\end{align}
%%%
Since there are no interactions between the different collinear sectors or the soft sector in \eq{LSCET}, we can factorize the forward matrix element into a product of several matrix elements.

First, for each final state collinear sector we get the vacuum expectation value of two collinear fields, with an insertion of $\delta[\w_i - \cE^0_i]$ between the fields, which restricts the collinear energy configuration to $\w_i$. Since the matrix element conserves four momentum, we can write it as
%%%
\begin{equation} \label{eq:J_def}
\mae{0}{ \phi(p_i') \, \delta[\w_i - \cE_i^0]\, \phi^\dagger(p_i) }{0}
= (2 \pi)^4 \delta^4(p_i' - p_i)\, J(p_i; \w_{i})
\,,\end{equation}
%%%
which defines the momentum-space jet function $J(p_i; \w_i)$. Integrating both sides over $p_i$, we obtain the equivalent definition in position space in terms of the standard fields $\phi_{n, \tp}(x)$,
%%%
\begin{equation}
J(p_i; \w_{i})
= \int\!\df^4 x \, e^{\img k_i \cdot x} \, \mae{0}{ \phi_{n_i, \tp_i}(x) \, \deltai{i} \, \phi^\dagger_{n_i, \tp_i}(0) }{0}
\,.\end{equation}
%%%
Momentum conservation implies that $J(p;\w)$ only has support for configurations $\w$ that satisfy $P[\w] = p$, where $P[\w]$ is defined in \eq{Pw}. One can also define a jet function $J(\w) = \int\!\df^4 p\, J(p;\w)$, which has support for any (physically allowed) $\w$, but we find it conceptually and notationally easier to keep the total momentum of $\w$ as an explicit separate argument.

At leading order in the power counting, any hadron in the initial state is bound by collinear interactions only, and thus does not interact with the soft sector. Hence, we can factor out the soft matrix element,
%%%
\begin{equation} \label{eq:S_def}
\mae{0}{ \cO_S^\dagger(k_s') \, \delta[\w_s - \cE_s^0]\, \cO_S(k_s) }{0}
= (2 \pi)^4 \delta^4(k_s' - k_s)\,(2\pi)^4 \delta(k_s - P[\w_s])\, S(\w_s)
\,,\end{equation}
%%%
which defines the soft function $S(\w_s)$, and we again used momentum conservation. Note that $\cO_S$ and $S$ depend on the $N+2$ collinear directions with respect to which the Wilson lines are defined, which is hidden in our notation. The soft function $S(\w_s)$ is defined with support for any physical $\w_s$, and the total soft momentum is given by $k_s = P[\w_s]$.

The remaining initial state matrix element can be written as
%%%
\begin{align}
\label{eq:I_def}
&\mae{\ini}{ \cO_I^\dagger (p_a', p_b')\, \deltai{a}\, \deltai{b}\, \cO_I (p_a, p_b)}{\ini}
\nn\\ & \qquad
= (2\pi)^4 \delta^4(p_a' - p_a)\,(2\pi)^4 \delta^4(p_b' - p_b)\, I(p_a, p_b; \w_a, \w_b)
\,,\end{align}
%%%
and defines the initial-state function $I(p_a, p_b; \w_a, \w_b)$. In writing \eq{I_def} we already used that the matrix element will factorize for the two collinear sectors, which allows us to write two separate momentum-conserving $\delta$ functions. As for the jet function, we choose to keep the momenta $p_{a,b}$ explicit in the definition of $I(p_a, p_b; \w_a, \w_b)$, so its support is restricted by momentum conservation to $P[\w_a+\w_b] = p_I - p_a - p_b$. For $e^+e^-$ collisions, the initial-state function reduces to a calculable leptonic matrix element, as discussed in \subsec{ee_generic}. For hadronic collisions, it can be reduced to parton distribution functions, but also allows one to treat the underlying event or beam remnants, as discussed in \subsec{pp_initial}.

Combining Eqs.~\eqref{eq:J_def}, \eqref{eq:S_def}, and \eqref{eq:I_def} with \eq{matching_step3}, we can perform the integrals over all primed momenta, and arrive at
%%%
\begin{align} \label{eq:dsigma_dw}
\frac{\delta \sigma}{\delta \w}
&= \frac{1}{2 p_I^2} \biggl( \prod_{i=1}^N \int\!\frac{\df^4 p_i}{(2 \pi)^4}\,\Df \w_i  \, J(p_i; \w_i) \biggr)
\int\!\frac{\df^4 p_a}{(2 \pi)^4}\, \frac{\df^4 p_b}{(2 \pi)^4}\, \Df \w_a\,
 \Df \w_b\, I(p_a, p_b; \w_a, \w_b)
\nn\\ & \quad \times
\abs{C_N(p_a, p_b; p_1, \dots , p_N)}^2
\int\!\Df\w_s\, S(\w_s)
\nn\\ & \quad \times
(2 \pi)^4 \delta^4\biggl(p_a + p_b - P[\w_s] - \sum_{i=1}^N p_i\biggr)\, \delta\biggl[\w - \w_s - \sum_{i=a,b,1}^N \w_i\biggr]
\,.\end{align}
%%%
As anticipated, the fully differential cross section $\delta \sigma / \delta \w$ is given by the product of a hard coefficient, $\abs{C_N}^2$, $N$ jet functions, $J(p_i; \w_i)$, an initial-state function, $I(p_a, p_b; \w_a, \w_b)$, and a soft function, $S(\w_s)$. Note that there are no power corrections to \eq{dsigma_dw} other than from higher-order SCET operators in the matching of QCD onto SCET and higher-order contributions to the Lagrangian, which could in principle be included systematically.
One should keep in mind that this factorization is purely academic at this point, because all ingredients depend on the precise energy configuration in each sector of the theory. The energy configurations are obviously very different for partonic and hadronic states, and therefore the functions $J$, $I$, and $S$ cannot be calculated perturbatively. One should think of them as fully exclusive functions.

The importance of \eq{dsigma_dw} lies in the fact that it establishes factorization for a generic $N$-jet like kinematic configuration. In our formalism, the question whether the cross section $\df\sigma/\df O$ for a particular observable factorizes is two-fold. First, a given value of the observable has to be dominated by factorizable kinematic configurations. If this is the case, one can immediately obtain a factorized form for $\df\sigma/\df O$ from \eq{dsigma_dw} via \eq{dsigma_dO2}. This means that \emph{any} jet observable (meaning any observable whose definition restricts it to $N$-jet configurations) is formally factorizable. The second, and more important, question then is whether one is able to determine the relevant functions, $J$, $I$, and $S$, for a given observable.

For sufficiently inclusive observables, the jet functions, $J(p_i; \w_i)$, will be smeared enough, \ie, integrated over $\w_i$ with a sufficiently smooth weight function, such that we can trust their perturbative calculation. Similarly, the soft function, $S(\w_s)$, and (for hadronic collisions) the initial-state function, $I(p_a, p_b; \w_a, \w_b)$, have to be smeared enough (integrated over $\w_s$ and $\w_{a,b}$) to reduce to nonperturbative functions that are universal between different processes. For such observables, one obtains a factorization formula in the more traditional sense, which allows for the perturbative calculation of all ingredients, with the exception of maybe a traditional soft function or initial state parton distribution functions.

To study the structure of the factorization for a specific observable, and obtain explicit definitions of the relevant jet, soft, and initial state functions, it is usually required to also expand the kinematics of the process, because \eq{dsigma_dw} still mixes momentum components with different scaling in SCET. In this way, one obtains a result that formally is fully leading order in the power counting.
As discussed above, with our choice of field labels the components $p_i^\mp = p_i^0 \pm \abs{\vec{p}_i}$ are defined with respect to the direction of the momentum $p_i$ itself, so $p_{i\perp} = 0$ and $\vn_i = \vp_i/\abs{\vp_i}$. Since $p_i^-\gg p_i^+$, to leading order the phase space in \eq{p_space} is
%%%
\begin{equation} \label{eq:p_space_exp}
\int\!\frac{\df^4 p_i}{(2 \pi)^4}
= \int\!\frac{\df p_i^-\, \df p_i^+\, \df \W_i}{(2 \pi)^4}\, \frac{(p_i^-)^2}{8}
\,.\end{equation}
%%%
Furthermore, expanding the momentum conserving $\delta$ function, we find
%%%
\begin{equation} \label{eq:momcons_exp}
\delta^4\biggl(p_a+p_b- P[\w_s] - \sum_{i=1}^N p_i\biggr ) = \delta^4\biggl(p_a^-\frac{n_a}{2} + p_b^- \frac{n_b}{2}- \sum_{i=1}^N p_i^-\frac{n_i}{2}\biggr)
\,.\end{equation}
%%%

Equation~\eqref{eq:dsigma_dw} together with \eqs{p_space_exp}{momcons_exp} provides the final factorized form of the fully differential cross section $\delta\sigma / \delta \w$ for $N$ jets, and is the main result of this paper. In the remaining part of the paper we will show how to use this result to understand the factorization properties of several observables. We will focus mostly on simple two-jet final states, for which the kinematics is simple enough to explicitly perform all phase space integrations analytically. All our examples, however, follow directly from our general result, and the extensions to more complicated final states should be obvious.

%%%%%%%%%%%%%%%%%%%%%%%%%%%%%%%%%%%%%%%%%%%%%%%%%%%%%%%%%%%%%%%%%%%%%%%%%%%%%%%%
\section{\boldmath $e^+ e^- \to 2$ Jets}
\label{sec:ee}
%%%%%%%%%%%%%%%%%%%%%%%%%%%%%%%%%%%%%%%%%%%%%%%%%%%%%%%%%%%%%%%%%%%%%%%%%%%%%%%%

In this section, we show how to apply the result in \eq{dsigma_dw} to the simplest case of two-jet events in $e^+ e^-$ collisions. The analysis simplifies considerably because of the absence of strongly interacting particles in the initial state, and due to the back-to-back nature of the jets and the corresponding need for only a single collinear direction, \eg, the thrust axis. We first give explicit definitions of the operators $\cO_I$, $\cO_F$, and $\cO_S$, including all relevant spin and color information, and then define all the ingredients in the factorized form of $\delta \sigma / \delta \w$. We then apply this generic formula to the special cases of event shape observables in the limit $e \to 1$ and to hemisphere jet masses, whose factorization is well understood~\cite{Korchemsky:1998ev, Korchemsky:1999kt, Korchemsky:2000kp, Belitsky:2001ij, Berger:2003iw}. Factorization for the former was proven using SCET in Refs.~\cite{Lee:2006nr, Bauer:2008dt} and for the latter in Ref.~\cite{Fleming:2007qr}, and we show how to reproduce these results. We then consider the factorization of generic observables defined for cone jet algorithms, and obtain the definition of the relevant cone jet functions and cone soft function. In SCET, cone jets were previously discussed in Refs.~\cite{Bauer:2002ie, Bauer:2003di, Trott:2006bk} using Sterman-Weinberg cones.

%===============================================================================
\subsection{Generic Expression}
\label{subsec:ee_generic}
%===============================================================================

For $e^+ e^- \to 2$ jets, including the full spin and color information, the three SCET operators entering the matching in \eq{matching_step2} are\footnote{We only give the result for an intermediate photon here, and include the $Z$ boson contribution later.}
%%%
\begin{align}
\cO_I^{\mu}(p_a, p_b) &= \bar{e}(-p_a) \gamma^\mu  e(p_b)
\,,\nn\\
\cO_{F\mu}^{cd}(p_1, p_2) &= \bar{\chi}^c(p_1)\gamma_\mu \chi^d(-p_2)
\,,\nn\\
\cO_S^{c d} (k_s)
&= \int\!\df^4 x\,e^{-\img k_s\cdot x}\, T\bigl[ Y_{n_1}^{\dagger\, ce}(x)\,Y_{n_2}^{ed}(x)\bigr]
\,, \end{align}
%%%
where $\chi^c(p) = [W \xi]^c(p)$ denotes a noninteracting collinear quark field of color $c$ and charge $e\,Q_f$ (where $f$ denotes the flavor), moving in the $\vec{p}/\vert\vec{p}\vert$ direction. Note that we are distinguishing particle and anti-particles by the sign of the momentum argument on the field. The soft Wilson lines along an outgoing collinear direction are%
\footnote{For a discussion of the different choices of boundary conditions for in- and outgoing Wilson lines see for example Ref.~\cite{Arnesen:2005nk}.}
%%%
\begin{equation}
\label{eq:Y_out}
Y_{n}^\dagger(x) = P\exp\biggl[\img g_s \int^{\infty}_0\!\df s \, n \cdot A_s(x+s\,n)\biggr]
\,,\end{equation}
%%%
where $P$ denotes path ordering. The Wilson coefficient at tree level is given by
%%%
\begin{equation}
C_2(p_a, p_b; p_1,p_2) = \frac{\img e^2\, Q_f}{2\,p_a\cdot p_b}\,[1 + \ord{\alpha_s}]
\,.\end{equation}
%%%

Since the initial state is not strongly interacting, the initial-state function in \eq{I_def}, including the average over initial spins, reduces to
%%%
\begin{align} \label{eq:I_leptonic}
&I^{\mu\nu}(p_a, p_b; \w_a, \w_b)
\nn\\ & \quad
= \frac{1}{4} \sum_{\rm spins} \int\!\frac{\df^4 p_a'}{(2\pi)^4}\,\frac{\df^4 p_b'}{(2\pi)^4}\,
\Mae{e^+e^-}{ \bar{e}(p_b')\gamma^\mu e(-p_a')\,\deltai{a}\,\deltai{b}\, \bar{e}(-p_a)\gamma^\nu  e(p_b)}{e^+e^-}
\nn\\ & \quad
= (2\pi)^4\delta(p_a - p_{e^+}) \,(2\pi)^4\delta(p_b - p_{e^-})\,\delta[\w_a]\,\delta[\w_b]\,L^{\mu\nu}
\,,\end{align}
%%%
where $p_{e^\pm}$ are the momenta of the incoming leptons and
%%%
\begin{equation}
L^{\mu\nu} = p_{e^-}^\mu p_{e^+}^\nu + p_{e^+}^\mu p_{e^-}^\nu - g^{\mu\nu}\,(p_{e^-}\mcdot p_{e^+})
\end{equation}
%%%
is the well-known leptonic tensor. Note that as we do not consider any initial state radiation from the incoming leptons, $\cE_{a,b}^0 = 0$ in \eq{I_leptonic}.

Using $\phi(p_1) = \chi_\alpha^c(p_1)$ and $\phi(p_2) = \bar\chi_\beta^d(-p_2)$ in \eq{J_def} (where $\alpha$, $\beta$ are spinor indices), the quark and anti-quark jet functions become (with the sum over spins implicit)
%%%
\begin{align} \label{eq:Jquark_nonsinglet}
J^{c'c}_{\alpha'\alpha}(p_1; \w_1)
&= \int\!\frac{\df^4 p_1'}{(2\pi)^4}\,\Mae{0}{\chi^{c'}_{\alpha'}(p_1')\, \deltai{1}\,
   \bar\chi^c_\alpha(p_1)}{0}
= \delta^{c'c} \Bigl(\frac{\dslash{n}_1}{2}\Bigr)_{\alpha'\alpha} J(p_1; \w_1)
\,,\\
\bar{J}^{d'd}_{\beta'\beta}(p_2; \w_2)
&= \int\!\frac{\df^4 p_1'}{(2\pi)^4}\,\Mae{0}{\bar\chi^{d'}_{\beta'}(-p_2')\, \deltai{2}\,
   \chi^d_\beta(-p_2)}{0}
= \delta^{d'd} \Bigl(\frac{\dslash{n}_2}{2}\Bigr)_{\beta\beta'} \bar{J}(p_2; \w_2)
\nn\,,\end{align}
%%%
where the spin-singlet and color-singlet jet functions are defined as
%%%
\begin{align} \label{eq:Jquark_singlet}
J(p_1; \w_1)
&= \frac{1}{4 N_c} \int\!\frac{\df^4 p_1'}{(2\pi)^4}\, \tr \Mae{0}{\bnslash_1 \chi(p_1') \, \deltai{1} \,
   \bar{\chi}(p_1)}{0}
\,,\nn\\
\bar{J}(p_2; \w_2)
&= \frac{1}{4 N_c} \int\!\frac{\df^4 p_2'}{(2\pi)^4}\, \tr \Mae{0}{\bar{\chi} (-p_2') \, \deltai{2}\,
   \bnslash_2 \chi(-p_2)}{0}
\,.\end{align}
%%%
Here, $\tr$ denotes the trace over spin and color indices and $N_c$ is the number of colors. At lowest order in perturbation theory, we have $J(p;\w) = \bar{J}(p;\w) = 2\pi\,\delta(p^+)\,\theta(p^-)\,\delta[\w(\W) - p^0\, \delta(\W-\W(\vp))]$.

From \eq{S_def}, the soft function is defined as
%%%
\begin{align}
S_{n_1n_2}^{d'c'cd}(\w_s) = \frac{1}{N_c} \int\!\frac{\df^4 k_s'}{(2\pi)^4}\,\frac{\df^4 k_s}{(2\pi)^4}\,
   \Mae{0}{\cO_S^{\dagger\, d'c'}(k_s')\, \deltai{s}\, \cO_S^{cd}(k_s)}{0}
\,,\end{align}
%%%
where we made explicit the dependence of $S$ on the directions $n_{1,2}$ of the Wilson lines in $\cO_S$, and
the factor $1/N_c$ is included by convention. Contracting with the trivial color structure of the jet functions in \eq{Jquark_nonsinglet}, we obtain the color-singlet soft function
%%%
\begin{equation}  \label{eq:Sqq_singlet}
S_{n_1n_2}(\w_s) = \delta^{c'c}\,\delta^{d'd}\, S_{n_1n_2}^{d'c'cd}(\w_s)
% \nn\\
= \frac{1}{N_c}\,\tr\Mae{0}{\overline{T}\bigl[(Y_{n_2}^{\dagger} Y_{n_1})(0)\bigr]\, \deltai{s}\,
T\bigl[ (Y_{n_1}^\dagger Y_{n_2})(0)\bigr]}{0}
\,.\end{equation}
%%%

Since the spin structure of the jet functions in \eq{Jquark_nonsinglet} factorizes, we can contract all vector and spinor indices,
%%%
\begin{equation}
L_{\mu\nu}\, \Bigl(\frac{\dslash{n}_1}{2}\Bigr)_{\alpha'\alpha} \gamma^\nu_{\alpha\beta}
\Bigl(\frac{\dslash{n}_2}{2}\Bigr)_{\beta\beta'} \gamma^\mu_{\beta'\alpha'}
= \Ecm^2(1 - \cos\theta_1\,\cos\theta_2)
\,,\end{equation}
%%%
where $\Ecm$ is the total energy and $\theta_{1,2}$ are the angles of $\vp_{1,2}$ with respect to the $e^+e^-$ beam axis in the center-of-mass frame. Thus, combining all pieces with \eq{dsigma_dw}, we find
%%%
\begin{align} \label{eq:dsigma_dw_eefull}
\frac{\delta \sigma}{\delta \w}
&= \frac{8\pi^2\alpha^2\, Q_f^2\, N_c}{\Ecm^4}
   \!\int\!\frac{\df^4 p_1}{(2 \pi)^4}\, \Df \w_1\, J(p_1; \w_{1})
   \!\int\!\frac{\df^4 p_2}{(2 \pi)^4}\,\Df \w_2  \, \bar{J}(p_2; \w_{2})
H_2(p_1, p_2)\, ( 1-\cos\theta_1\cos\theta_2)
\nn\\ & \quad \times
\int\!\Df\w_s\, S_{n_1n_2}(\w_s)\,
(2 \pi)^4 \delta^4(p_{e^+} + p_{e^-} - P[\w_s] - p_1 - p_2)\, \delta[\w - \w_1 - \w_2 - \w_s]
\,,\end{align}
%%%
where the hard coefficient $H_2(p_1, p_2) = 1 + \ord{\alpha_s}$ is defined by
%%%
\begin{equation} \label{eq:H2_def}
\abs{C_2(p_{e^+},p_{e^-};p_1, p_2)}^2
= \Bigl(\frac{4\pi\alpha\, Q_f}{\Ecm^2}\Bigr)^2 H_2(p_1, p_2)
\,.\end{equation}
%%%
Equation~\eqref{eq:dsigma_dw_eefull} specializes \eq{dsigma_dw} to generic $2$-jet configurations $\w$ in $e^+e^-$ collisions.

Next, we expand the kinematics. Using \eq{momcons_exp}, the momentum conserving $\delta$ function becomes
%%%
\begin{align}
&\delta^4\Bigl(p_{e^+}^-\frac{n_{e^+}}{2} + p_{e^-}^- \frac{n_{e^-}}{2} - p_1^-\frac{n_1}{2} - p_2^-\frac{n_2}{2}\Bigr)
\nn\\ &\qquad
= \frac{8}{\Ecm^2}\, \delta(p_1^- - \Ecm)\, \delta(p_2^- - \Ecm)\, \delta(\cos{\theta}_1 + \cos{\theta}_2)\, \delta(\phi_1 - \phi_2 - \pi)
\,,\end{align}
%%%
where as before in the center-of-mass frame $p_I = p_{e^+} + p_{e^-} = (\Ecm, \vec{0})$. The $\delta$ functions allow us to perform the $p_1^-$, $p_2^-$, and $\W_2$ integrations in \eq{dsigma_dw_eefull}, and imply that $p_1$ and $p_2$ are back-to-back, as expected for two-jet events. In particular, $\vn_1 = -\vn_2$, so we can write $p_{1,2}$ in terms of the components $(p^+, p^-, \vn)$ as
%%%
\begin{equation} \label{eq:p12_exp}
p_1 = \bigl(p_1^+, \Ecm, \vn(\W) \bigr)
\,,\qquad
p_2 = \bigl(p_2^+, \Ecm, -\vn(\W) \bigr)
\,,\end{equation}
%%%
where $\W = (\theta, \phi) \equiv \W_1$ describes the orientation of the momenta relative to the beam axis. We also write $S_{n_1n_2} \equiv S_{\vn(\W)}$. Similar to the Wilson coefficient $C_2$, the hard coefficient $H_2(p_1,p_2)$ does not depend on the small momentum components $p_{1,2}^+$. Since $p_i^- = \Ecm$, we define $H_2(\Ecm) \equiv H_2(p_1, p_2)$.
Combining everything with \eq{dsigma_dw_eefull}, using \eq{p_space_exp}, and writing the momenta in terms of their components, we have
%%%
\begin{align} \label{eq:dsigma_dw_eeexp}
\frac{\delta \sigma}{\delta \w}
&= H_2(\Ecm) \int\!\frac{\df \W}{2 \pi}
   \int\!\frac{\df p_1^+}{2 \pi}\, \Df\w_1\, J(p_1^+, \Ecm, \vn(\W); \w_1)
   \int\!\frac{\df p_2^+}{2 \pi}\, \Df\w_2\, \bar{J}(p_2^+, \Ecm, -\vn(\W); \w_2)
\nn\\ & \quad\times
   \frac{\df \sigma_0 }{\df \cos\theta} \int\!\Df\w_s \, S_{\vn(\W)}(\w_s)\,
   \delta[\w - \w_1 - \w_2 - \w_s]
\,,\end{align}
%%%
where
%%%
\begin{equation}
\frac{\df \sigma_0 }{\df\cos\theta}  = \frac{\pi \alpha^2}{2 \Ecm^2}\,N_c\, Q_f^2\, (1+\cos^2\theta)
\end{equation}
%%%
is the Born differential cross section. The exchange of a $Z$ boson can be included by using
%%%
\begin{align}
\frac{\df \sigma_0 }{\df\cos\theta}
&= \frac{\pi \alpha^2}{2 \Ecm^2}\,N_c\, \biggl\{
   \biggl[ Q_f^2 - \frac{2\,v_e v_f Q_f}{1- m_Z^2/\Ecm^2} +
   \frac{(v_e^2+a_e^2)(v_f^2+a_f^2)}{(1-m_Z^2/\Ecm^2)^2} \biggr] (1+\cos^2\theta)
\nn\\ & \quad
+ \biggl[ \frac{4\, a_e a_f Q_f^2}{1-m_Z^2/\Ecm^2} - \frac{8\,v_e a_e v_f a_f }{(1-m_Z^2/\Ecm^2)^2} \biggr] \cos\theta \biggr\}
\,,\end{align}
%%%
where $v_{e,f}$ and $a_{e,f}$ are the standard vector and axial couplings to the $Z$.

Equation~\eqref{eq:dsigma_dw_eeexp} is the penultimate formula for generic observables in $e^+ e^- \to 2$ jet events. Each of the ingredients in \eq{dsigma_dw_eeexp} is a completely exclusive object that depends on the energy distribution of the individual partons. The details of how to integrate over the energy configurations to arrive at perturbative jet functions and a universal soft function depend on the observable in question, but since all observable independent simplifications have been done, a wide range of factorization theorems can now be obtained with relative ease. We illustrate this with a few examples in the following subsections.

%===============================================================================
\subsection{\boldmath Event Shapes in the Limit $e \to 1$}
%===============================================================================

Combining \eqs{dsigma_dO2}{fe_wdef}, the differential cross section in some event shape $e$ is
%%%
\begin{equation}
\frac{\df\sigma}{\df e}
= \int\!\df\vt \int\!\Df\w\,\frac{\delta\sigma}{\delta\w}\,\delta(\vt - \vt[\w])\,\delta(e - f_e[\w;\vt])
\,.\end{equation}
%%%
For $e \to 1$, the final state is dominated by two highly collimated jets, and hence, we can use the result for $\delta\sigma/\delta\w$ in \eq{dsigma_dw_eeexp}. The integration over $\w$ is trivial and sets $\w = \w_1 + \w_2 + \w_s$. Since $\w_{1,2}$ describe collinear energy configurations along $\pm\vn(\W)$, we have
$\vt[\w_1 + \w_{2} + \w_s] = \vn + \ord{\lambda^2}$~\cite{Bauer:2008dt}. This allows us to integrate over $\vt$,
%%%
\begin{align} \label{eq:dsigma_de}
\frac{\df\sigma}{\df e}
&= H_2(\Ecm)\int\!\frac{\df \W}{2 \pi}
   \int\!\frac{\df p_1^+}{2 \pi}\, \Df\w_1\, J(p_1^+, \Ecm, \vn(\W); \w_1)
   \int\!\frac{\df p_2^+}{2 \pi}\, \Df\w_2\, \bar{J}(p_2^+, \Ecm, -\vn(\W); \w_2)
\nn\\ & \quad\times
   \frac{\df \sigma_0 }{\df \cos\theta} \int\!\Df\w_s \, S_{\vn(\W)}(\w_s)\,
   \delta(e - f_e[\w_1 + \w_2 + \w_s; \vn(\W)])
\,.\end{align}
%%%
From \eq{fe_wtdef}, we see that $f_e[\w;\vn]$ is linear in $\w$, from which it follows that we can write $f_e[\w_1 +\w_2 + \w_s; \vn] = f_e[\w_1; \vn] + f_e[\w_2; \vn] + f_e[\w_s; \vn]$. This implies
%%%
\begin{align} \label{eq:few_separation}
\delta (e - f_e[\w_1 +\w_2 + \w_s; \vn])
&= \int\!\df e_1\, \df e_2\, \df e_s  \, \delta(e - e_1 - e_2 - e_s)
\nn\\ & \qquad \times
   \delta(e_1 - f_e[\w_1; \vn]) \, \delta(e_{2} - f_e[\w_{2}; \vn]) \,\delta(e_s - f_e[\w_s; \vn])
\,,\end{align}
%%%
which separates the $\w$ dependencies in \eq{dsigma_de}. We stress that this is not a requirement for the factorization of $\df\sigma/\df e$, as demonstrated by \eq{dsigma_de}. In fact, the full event-shape functional $f_e[\w] = f_e[\w, \vt[\w]]$ is not linear in $\w$ and does not obey a similar separation, because $\vt[\w]$ is not linear in $\w$. The crucial ingredient for the factorization is the linearity of the energy-momentum tensor and the resulting separation of the energy flow operator in \eqs{Eflowlinearity}{Eflowfactorization}.
However, without \eq{few_separation} the jet and soft functions depend on the full energy distributions $\w_i$, and are therefore neither perturbatively calculable, nor universal enough to be extracted from data. The important point about \eq{few_separation} is that it allows us to perform the $\w$ integrations in \eq{dsigma_de}, and to define inclusive event-shape jet and soft functions
%%%
\begin{align} \label{eq:JSe_def}
J(e_1)
&= \int\!\frac{\df p_1^+}{2 \pi}\,\Df \w_1\, J(p^+_1, \Ecm,\vn; \w_1)\, \delta(e_1 - f_e[\w_1; \vn])
\,,\nn\\
\bar{J}(e_2)
&= \int\!\frac{\df p_2^+}{2 \pi}\,\Df \w_2\, \bar{J}(p^+_2, \Ecm, \vn; \w_2)\, \delta(e_2 - f_e[\w_2; \vn])
\,,\nn\\
S(e_s) &= \int\!\Df \w_s\, S_\vn(\w_s)\, \delta(e_s - f_e[\w_s; \vn])
\,.\end{align}
%%%
With the definitions in \eqs{Jquark_singlet}{Sqq_singlet}, these are identical to the definitions given in Ref.~\cite{Bauer:2008dt}. For $\bar{J}(e_2)$ we used that $f_e[\w,\vn] = f_e[\w,-\vn]$ because the sign of the thrust vector is irrelevant. By rotational invariance, after integrating over $\w_{1,2}$, the jet functions $J(e_1)$, $\bar{J}(e_2)$, do not depend on the value of $\vn$ on the right-hand side. This would not be true if the thrust axis $\vn$ in $f_e[\w;\vn]$ would be different from the momentum direction $\vn$ in $J(p^+,\Ecm,\vn;\w)$. Similarly, after integrating over $\w_s$, the soft function $S(e_s)$ is independent of $\vn$, because the direction of the Wilson lines in $S_\vn(\w_s)$ coincides with the thrust axis. Thus, using \eq{JSe_def} and integrating over $\W$, we obtain the final result
%%%
\begin{align}
\label{eq:dsigma_de_final}
\frac{\df \sigma}{\df e}
&= H_2(\Ecm)\,\sigma_0
\int\!\df e_1 \, \df e_2 \, \df e_s \, J(e_1)\, \bar{J}(e_2)  \, S(e_s) \, \delta (e - e_1 - e_2 - e_s)
\,.\end{align}
where $\sigma_0 = \int\!\df\cos\theta\,\df\sigma_0/\df\cos\theta$ is the total Born cross section. Equation~\eqref{eq:dsigma_de_final} agrees with the result of Ref.~\cite{Bauer:2008dt}.

%~~~~~~~~~~~~~~~~~~~~~~~~~~~~~~~~~~~~~~~~~~~~~~~~~~~~~~~~~~~~~~~~~~~~~~~~~~~~~~~
\subsection{Double Differential Hemisphere Mass Distribution}
%~~~~~~~~~~~~~~~~~~~~~~~~~~~~~~~~~~~~~~~~~~~~~~~~~~~~~~~~~~~~~~~~~~~~~~~~~~~~~~~

Combining \eqs{dsigma_dO2}{fO_jetP}, the double differential hemisphere mass distribution is
%%%
\begin{equation} \label{eq:dsigma_dM1dM2_def}
\frac{\df^2 \sigma}{\df M_1^2\, \df M_2^2}
= \int\!\df^4 P_1\int\!\df^4 P_2 \, \frac{\df^2\sigma}{\df^4P_1\,\df^4P_2}\, \delta(M_1^2 - P_1^2)\, \delta(M_2^2 - P_2^2)
\,,\end{equation}
%%%
where the cross section differential in the total momentum of each jet for the hemisphere jet algorithm is
%%%
\begin{equation} \label{eq:dsigma_dPjet}
\frac{\df^2 \sigma}{\df^4 P_1\, \df^4 P_2}
= \int\!\df\vt \int\!\Df\w\,\frac{\delta\sigma}{\delta\w}\,\delta(\vt - \vt[\w])
\,\delta^4(P_1 - P[\cJ_1[\w;\vt]])\,\delta^4(P_2 - P[\cJ_2[\w;\vt]])
\,.\end{equation}
%%%
Here, $P[\w]$ is given in \eq{Pw} and $\cJ_i[\w;\vt]$ in \eq{hemi_J}. Combining these we have
%%%
\begin{align} \label{eq:Piw_hemi}
P_{\hemi\,1}^\mu[\w;\vt] &\equiv P^\mu[\cJ_1[\w;\vt]]
= \int \! \df\W \, n^\mu(\W)\, \w(\W)\, \theta(0 \leq \theta_\vt < \pi/2)
\,,\nn\\
P_{\hemi\,2}^\mu[\w;\vt] &\equiv P^\mu [\cJ_2[\w;\vt]]
= \int \! \df\W \, n^\mu(\W)\, \w(\W)\, \theta(\pi/2 \leq \theta_\vt \leq \pi)
\,.\end{align}
%%%
In general, \eq{dsigma_dPjet} will receive contributions from final states containing several distinct collinear momenta, corresponding to SCET operators with $N \geq 2$. However, if the final states are restricted to the kinematic region of small hemisphere invariant masses $M_i^2 = P_i^2 \sim \ord{\lambda^2 \Ecm^2}$, corresponding to two collimated jets, the operator with $N=2$ collinear directions gives the dominant contribution, with the corrections suppressed by powers of $\lambda$. Thus, we can apply the result for $\delta\sigma/\delta\w$ in \eq{dsigma_dw_eeexp} in this region.

The integral over $\vt$ can be performed as in the previous subsection, which sets $\vt = \vn + \ord{\lambda^2}$. As \eq{Piw_hemi} is linear in $\w$, we then have
%%%
\begin{equation} \label{eq:Piscaling}
P_{\hemi\,i}[\w;\vn] = P_{\hemi\,i}[\w_1+\w_2+\w_s;\vn] = P_{\hemi\,i}[\w_1;\vn] + P_{\hemi\,i}[\w_2;\vn] + P_{\hemi\,i}[\w_s;\vn]
\,.\end{equation}
%%%
To understand the size of $P_{\hemi\,i}[\w_j]$  for $i = j$ and $i \neq j$, we need to think about states in SCET in some more detail. Since the direction $n$ labelling the collinear fields in SCET is a conserved quantum number, there exists a basis for the physical states which have a fixed value of $n$ as well. This implies that for a given SCET state with momentum $p$, one has to identify the value of the direction $n$ as well. Of course, to have the same final states as in full QCD, one needs $\sum_n \ket{p,n}^\mathrm{SCET} = \ket{p}^\mathrm{QCD}$, \ie~one has to be careful not to double count the physical states. Certainly, a convenient choice is to define the SCET states such that for every momentum $p$ there is only a single value $n$. For our problem, the simplest choice is to assign the label $n_1 = (1,\vn)$ to all states with momentum in hemisphere 1, and $n_2 = (1,-\vn)$ to all states in hemisphere 2. This choice implies
%%%
\begin{equation}
P_{\hemi\,i}[\w_{j\neq i};\vn] = 0
\qquad\text{and} \qquad
P_{\hemi\,i}[\w_i;\vn] = p_i
\,,\end{equation}
%%%
where $p_i = P[\w_i]$ is the total momentum of $\w_i$, \ie, the momentum in $J(p_i;\w_i)$. The power counting of SCET implies that $\ell_i = P_{\hemi\, i}[\w_s] \sim \lambda^2\Ecm^2$, where $\ell_i$ can be interpreted as the total soft momentum in each hemisphere. Thus, using \eq{p12_exp} we can expand
%%%
\begin{equation}
M_i^2 = P_i^2 = p_i^2 + 2p_i\cdot\ell_i + \ell_i^2 = \Ecm(p_i^+ + n_i\cdot \ell_i) + \ord{\lambda^4\Ecm^2}
\,.\end{equation}
%%%

Since our observables $M_i^2$ only depend on $p_i^+$ and $n_i\cdot\ell_i$, we can do the remaining integrations in \eq{dsigma_dw_eeexp}, and define the corresponding jet and soft functions
%%%
\begin{align}
J(\Ecm p_1^+) &= \frac{1}{2\pi\Ecm}\int \! \Df \w_1 \, J(p_1^+, \Ecm, \vn;\w_1)
\,,\nn\\
\bar{J}(\Ecm p_2^+) &= \frac{1}{2\pi\Ecm}\int \! \Df \w_2 \, \bar{J}(p_2^+, \Ecm, \vn;\w_2)
\,,\nn\\
S_\hemi(\ell_1^+,\ell_2^+)
&=\int\!\Df \w_s\, S_\vn(\w_s)\,
   \delta(\ell_1^+ - n_1 \mcdot P_{\hemi\,1}[\w_s;\vn])\, \delta(\ell_2^+ - n_2 \mcdot P_{\hemi\,2}[\w_s;\vn])
\,.\end{align}
%%%
Again, after integrating over $\w_{1,2,s}$, the jet functions, $J(p_1^2)$ and $\bar{J}(p_2^2)$, and the hemisphere soft function, $S_\hemi(\ell_1^+,\ell_2^+)$, do not depend on $\vn$ due to rotational invariance. The SCET hemisphere soft function has been discussed previously in Refs.~\cite{Fleming:2007qr, Hoang:2007vb, Fleming:2007xt, Hoang:2008fs}. The above definition provides an operator definition of $S_\hemi(\ell_1^+,\ell_2^+)$ in SCET, and is equivalent to the definition in Ref.~\cite{Korchemsky:1999kt}. Putting everything together, we obtain for the double differential hemisphere mass distribution
%%%
\begin{equation}
\label{eq:dsigma_dM1dM2}
\frac{\df^2 \sigma}{\df M_1^2\, \df M_2^2}
= H_2(\Ecm)\, \sigma_0 \int\!\df\ell_1^+\,\df\ell_2^+\, J(M_1^2 - \Ecm\ell_1^+)\,\bar{J}(M_2^2 - \Ecm\ell_2^+)\, S_\hemi(\ell_1^+,\ell_2^+)\,
\,,\end{equation}
%%%
which agrees with the massless limit of the result in Ref.~\cite{Fleming:2007qr}.

%~~~~~~~~~~~~~~~~~~~~~~~~~~~~~~~~~~~~~~~~~~~~~~~~~~~~~~~~~~~~~~~~~~~~~~~~~~~~~~~
\subsection{Two-Jet Cone Algorithms}
\label{subsec:TwoJetCone}
%~~~~~~~~~~~~~~~~~~~~~~~~~~~~~~~~~~~~~~~~~~~~~~~~~~~~~~~~~~~~~~~~~~~~~~~~~~~~~~~

As the last example in this section, we consider the cross section for two-jet final states obtained from an infrared-safe cone jet algorithm. Since the discussion follows closely that of the previous two subsection, we keep it short, mainly highlighting the differences.
Combining \eqs{dsigma_dO2}{fO_jetP} with $\cN = 2$ we have
%%%
\begin{align} \label{eq:dsigma_dO_2cones_def}
\frac{\df \sigma}{\df O}
&= \int\!\df\vj_1\,\df\vj_2 \int \!\Df \w \, \frac{\delta\sigma}{\delta\w}\,
   \delta(\vj_1 - \vj_1[\w])\,\delta(\vj_2 - \vj_2[\w])
\nn\\ & \quad \times
   \int\!\df^4 P_1\, \delta^4 (P_1 - P_\cone[\w;\vj_1])
   \int\!\df^4 P_2\, \delta^4 (P_2 - P_\cone[\w;\vj_2]) \,\delta(O-g_O(P_1,P_2))
\,,\end{align}
%%%
where $\vj_i[\w]$ denotes the $i$-th jet direction returned by $\vj[\w]$, and the functionals for the total jet momenta are now defined for example using \eq{cone_J}
%%%
\begin{align} \label{eq:Piw_cone}
P_\cone^\mu[\w;\vj_i] &\equiv P^\mu[\cJ_i[\w;\vec{j}]] =  \int \! \df\W \, n^\mu(\W)\, \w(\W)\, \theta(R - \theta_{\vj_i})
\,.\end{align}
%%%
As before, \eq{dsigma_dO_2cones_def} receives in general contributions from operators with $N \geq 2$. However, if the final state is restricted to two jets with small invariant masses, $M_i^2 = P_i^2 \sim \ord{\lambda^2 \Ecm^2}$, the result for $\delta\sigma/\delta\w$ in \eq{dsigma_dw_eeexp} for $N=2$ gives the dominant contribution, with corrections suppressed by powers of $\lambda$. The restriction on the kinematics of the final state is now provided by the jet algorithm, or by the combination of jet algorithm and observable.

For a good jet algorithm, the result of $\vj[\w]$ should not depend on $\w_s$ up to power corrections. This is equivalent to the requirement that the jet algorithm should not be infrared sensitive. Furthermore, since $\w_i$ describes a collinear energy configuration along $\vn_i$, by a similar argument as in the case of thrust, up to power corrections, the direction of the jets is aligned with the direction of the collinear fields. Therefore,
%%%
\begin{equation} \label{eq:vjcone_wi}
\vj_i[\w_1 + \w_2 + \w_s] =  \vn_i + \ord{\lambda^k}
\,.\end{equation}
%%%
The power of $k$ depends on the details of the algorithm, \eg\ for the hemisphere jet algorithm, where $\vj$ is the thrust axis, we had $k = 2$.

To define the states in SCET, we assign the label $n_i$ to states with momentum lying in the $i$-th cone, so there is again no overlap between states with the same momentum but different $n$ inside the cones. The precise definition of states with momentum outside any of the cones is not important at this point. With this definition, $P_\cone[\w_{j\neq i}; \vn_i] = 0$, and since \eq{Piw_cone} is linear in $\w$, we have
%%%
\begin{equation} \label{eq:Piscaling_cone}
P_\cone[\w_1 + \w_2 + \w_s; \vn_i] = P_\cone[\w_i;\vn_i] + P_\cone[\w_s; \vn_i] \equiv q_i + \ell_i
\,,\end{equation}
%%%
where $q_i = P_\cone[\w_i;\vn_i]$ and $\ell_i = P_\cone[\w_s;\vn_i]$ are the total collinear and soft momentum in each cone. Equation~\eqref{eq:vjcone_wi} implies that $q_i + \ell_i$ are aligned along $\vn_i$ up to power corrections. In addition, note that $q_i^\pm \equiv q_i^\pm(R)$ is a function of the cone size $R$ (and the used jet algorithm). For $R = \pi$ the cones become hemispheres, and thus $q_i^\pm(\pi) = p_i^\pm$, while at lowest order in perturbation theory, $q_i^\pm(R) = p_i^\pm - \ord{\alpha_s}$. Thus, for large enough $R$, $q_i^\pm/p_i^\pm \sim 1$ with the corrections calculable in perturbation theory. (Generically, we expect the perturbative corrections to contain logarithms of $\pi/R$. Similar phase space logarithms have been studied for the case of Sterman-Weinberg jets in Ref.~\cite{Trott:2006bk}.) Hence, as $q_i^\pm$ obeys the same power counting as $p_i^\pm$ for reasonable $R$, any observable that does not vanish at leading order in the SCET power counting can be written as
%%%
\begin{equation} \label{eq:gO_cone}
g_O(P_1,P_2) \equiv g_O(q_1^+ + \ell_1^+, q_1^-, q_2^+ + \ell_2^+, q_2^-, \vn) + \ord{\lambda^m}
\,,\end{equation}
%%%
where, $m$ is not necessarily the same as $k$ and also depends on the observable.

Since \eq{gO_cone} only depends on $q_i^\pm$, the result for $\df\sigma/\df O$ can be expressed in terms of the cone jet and soft functions
%%%
\begin{align} \label{eq:JS_cone}
J_\cone(q_1^+,q_1^-)
&= \int \!\frac{\df p_1^+}{2 \pi}\,\Df \w_1\, J(p_1^+, \Ecm,\vn; \w_1)\, \delta(q_1^+ - n_1\mcdot P_\cone[\w_1; \vn])\, \delta(q_1^- - n_2\mcdot P_\cone[\w_1; \vn])
\,,\nn\\
\bar{J}_\cone(q_2^+, q_2^-) &= \int \!\frac{\df p_2^+}{2 \pi}\,\Df\w_2\, \bar{J}(p_2^+, \Ecm, \vn; \w_2)\, \delta(q_2^+ - n_1\mcdot P_\cone[\w_2; \vn])\, \delta(q_2^- - n_2\mcdot P_\cone[\w_2; \vn])
\,,\nn\\
S_\cone(\ell_1^+, \ell_2^+)
&= \int \!\Df\w_s\, S_\vn(\w_s)\, \delta(\ell_1^+-n_1\mcdot P_\cone[\w_s;\vn])\, \delta^4(\ell_2^+-n_2\mcdot P_\cone[\w_s;-\vn])
\,,\end{align}
%%%
where as before $n_{1,2} = (1,\pm\vn)$, and the functions on the left-hand side do not depend on $\vn$.
Combining \eq{dsigma_dw_eeexp} with \eq{dsigma_dO_2cones_def} and using the above definitions, we obtain the
final result for the factorized differential cross section
%%%
\begin{align} \label{eq:dsigma_dO_2cones}
\frac{\df \sigma}{\df O}
&= H_2(\Ecm)\int\!\frac{\df\W}{2\pi}\,\frac{\df\sigma_0}{\df\cos\theta}
   \int\!\df q_1^+\, \df q_1^- \, J_\cone(q_1^+,q_1^-)
   \int\!\df q_2^+\, \df q_2^- \, \bar{J}_\cone(q_2^+,q_2^-)
\nn\\ &\quad\times
   \int\!\df\ell_1^+\,\df\ell_2^+\,S_\cone(\ell_1^+, \ell_2^+)\,
   \delta\bigl(O - g_O(q_1^+ + \ell_1^+,q_1^-, q_2^+ + \ell_2^+,q_2^-, \vn(\W))\bigr)
\,.\end{align}
%%%
To our knowledge, factorization for jet distributions has not received much attention in the literature (however, see Refs.~\cite{Trott:2006bk, Kidonakis:1998bk}), and this is the first time any factorization theorem for jet observables based on jet algorithms has been proven in the framework of SCET.

For many observables, such as the transverse momentum distribution, the dependence on the soft momenta and the small components $q_i^+$ is power suppressed, which allows us to integrate over these to obtain
%%%
\begin{align}
\frac{\df \sigma}{\df O} = & \,H_2(\Ecm) \int \!\frac{\df \W}{2 \pi}\, \frac{\df \sigma_0}{\df \cos \theta} \int\!\df q_1^- \, J_\cone(q_1^-)\int \df q_2^- \, \bar{J}_\cone(q_2^-)\, S_\cone \, \delta\bigl(O- g_O(q_1^-,q_2^-,\vn(\W))\bigr)
\,,\end{align}
%%%
where $\int \df l^+ \df l^- \, S_\cone (l^+, l^-) \equiv S_\cone$ is perturbatively calculable up to small power corrections and we defined
%%%
\begin{equation} \label{eq:Jcone_integrated}
J_\cone(q_1^-) = \int \! \df q_1^+\, J_\cone(q_1^+,q_1^-)
\,, \qquad
\bar{J}_\cone(q_2^-) = \int \! \df q_2^+\, \bar{J}_\cone(q_2^+,q_2^-)
\,.\end{equation}
%%%

%%%%%%%%%%%%%%%%%%%%%%%%%%%%%%%%%%%%%%%%%%%%%%%%%%%%%%%%%%%%%%%%%%%%%%%%%%%%%%%%
\section{Towards \boldmath $pp \to 2$ Jets}
\label{sec:pp}
%%%%%%%%%%%%%%%%%%%%%%%%%%%%%%%%%%%%%%%%%%%%%%%%%%%%%%%%%%%%%%%%%%%%%%%%%%%%%%%%

In the previous section we have focused on two-jet production in $e^+ e^-$ collisions. In this section, we extend these results to include hadrons in the initial state. Jet production in hadronic collisions is in several ways more complicated than for $e^+e^-$ collisions. First, there are several different partonic processes contributing to $pp \to 2$ jets. Second, the operators describing the short distance process now contain strongly interacting particles for both initial and final states, giving rise to a more involved color and Dirac structure. Finally, there are several additional matrix elements required to describe the long distance physics. These are the parton distribution function describing how the initial state partons are distributed inside the incoming proton, as well as new soft functions.

In this paper, we will only consider the simplest partonic process $q q' \to qq'$, and work only to tree level in the matching from QCD onto SCET. This simplifies the discussion dramatically, since only a single operator contributes at this order. Furthermore, due to the absence of gluons in the initial or final state, the only additional nonperturbative ingredients are the parton distributions of finding a quark inside the proton and the new soft function. The complete analysis of $pp \to 2$ jets is considerably more involved and will be discussed elsewhere~\cite{inprogress}.

%===============================================================================
\subsection{Matching onto SCET at Tree Level}
\label{subsec:pp_matching}
%===============================================================================

At tree level, only a single operator is required in SCET to describe the partonic process $qq' \to qq'$, schematically
%%%
\begin{equation}
\label{eq:opqqqq}
\cO(p_a,p_b; p_1,p_2) = \Cqq(p_a, p_b; p_1, p_2) \, \cO_I(p_a,p_b)\, \cO_F(p_1,p_2)\, \cO_S(k_s)
\,,\end{equation}
%%%
where we define the Wilson coefficient $C_4$ to contain all the kinematic and Dirac factors. The operators $\cO_I$, $\cO_F$, and $\cO_S$ are defined as
%%%
\begin{align}
\cO_{I\,\alpha\beta}^{cd} (p_a,p_b) &= \chi^{c}_{\alpha}(p_a) \, \chi^{d}_{\beta}(p_b)
\,, \nn\\
\cO_{F\,\gamma\delta}^{ef}(p_1,p_2) &=  \bar \chi^{e}_{\gamma}(p_1) \, \bar \chi^{f}_{\delta}(p_2)
\,,\nn\\
\cO_S^{ec\,fd}(k_s) &=
\int\!\df^4 x \,e^{-\img k_s \cdot x} \,
T\bigl[(Y_{n_1}^\dagger T^A\Yin_{n_a})^{ec}(x)\,(Y_{n_2}^\dagger T^A\Yin_{n_b})^{fd}(x) \bigr]
\,,\end{align}
%%%
where subscripts denote spinor and superscripts color indices.
The Wilson lines for the outgoing fields are defined as in \eq{Y_out}, while for the incoming fields they are
%%%
\begin{equation}
\Yin_{n}(x) = P\, \exp\biggl[ \img g_s \int_{-\infty}^0\!\df s \, n \cdot A_s(x+sn)\biggr]
\,.\end{equation}
%%%
The Wilson coefficient is given by
%%%
\begin{align} \label{eq:C4}
\Cqq(p_a, p_b; p_1, p_2) &= \frac{\img g_s^2}{\hat t}\, (\gamma^\mu)_{\gamma \alpha}\, (\gamma_\mu)_{\delta \beta}
\,,\end{align}
%%%
where we stress again that we are only working to tree level in the matching.
The variable $\hat t$ is one of the usual Mandelstam variables defined in terms of the partonic momenta
%%%
\begin{equation}
\hat s = (p_a + p_b)^2 \,, \qquad \hat t = (p_a - p_1)^2 \,, \qquad \hat u = (p_a - p_2)^2\,.
\end{equation}
%%%

%===============================================================================
\subsection{New Nonperturbative Matrix Elements}
\label{subsec:pp_initial}
%===============================================================================

There are two sources of additional matrix elements which cannot be calculated perturbatively. First, the operator $\cO_I$ now includes strongly interacting degrees of freedom, and the matrix elements involving the initial state protons are no longer calculable. Second, the soft operator contains four Wilson lines, rather than just two as for $e^+ e^-$ collisions. This implies that a new  soft function is required. In this section we define all required nonperturbative matrix elements needed for the process $pp \to 2$ jets via the partonic process $qq' \to qq'$.

Since the initial state hadrons are moving along different light cones, they are described by two sets of SCET Lagrangians which do not interact with each another. Therefore, the physics of the two initial states completely factorizes, in the same way as the final state partons in different directions factorize from one another, and we can write $\ket{I} = \ket{P_a}\,\ket{P_b}$ and $\cO_I(p_a,p_b) = \cO_I^a(p_a) \cO_I^b(p_b)$, such that we can factorize the initial state matrix element as
%%%
\begin{align}
&\mae{I}{ \cO^\dagger_I(p'_a,p'_b)\, \deltai{a}\, \deltai{b}\, \cO_I(p_a,p_b) }{I}
\nn\\ &\qquad
= \mae{P_a}{ \cO^{a\,\dagger}_I(p'_a)\, \deltai{a}\, \cO^a_I(p_a) }{P_a}\,
  \mae{P_b}{ \cO^{b\,\dagger}_I(p'_b)\, \deltai{b}\, \cO^b_I(p_b) }{P_b}
\,.\end{align}
%%%
For the case considered here, the operators $\cO_I^a$ and $\cO_I^b$ contain just a single quark field, $(\cO_I^a)^{c}_{\alpha} = \chi^{c}_{\alpha}(p_a)$ and $(\cO_I^b)^d_{\beta} = \chi^d_\beta(p_b)$. The resulting matrix elements define the parton distribution functions to find the quarks $q$ and $q'$ in the initial protons $P_{a,b}$.
%%%
\begin{equation}
\label{eq:generalqpdf}
\int\!\frac{\df^4 p_a'}{(2\pi)^4} \Mae{P_a}{ \bar \chi^{c'}_{ \alpha'}(p'_a)\, \deltai{a}\,  \chi^{c}_{\alpha}(p_a) }{P_a}
 = \frac{1}{2N_c}\,\delta^{c'c}\,\Bigl(\frac{\nslash_{a}}{2}\Bigr)_{ \alpha  \alpha'}\, f_{q/P}(p_a;\w_a)
\,,\end{equation}
%%%
and similarly for $f_{q'/P}(p_b;\w_b)$.
(Note that since we are distinguishing particles and antiparticles by the sign of their momentum, there is no anti-quark distribution on the right-hand side.) Combining these results gives the initial state function
%%%
\begin{equation}
I^{c'c\,d'd}_{\alpha'\alpha\,\beta'\beta}(p_a,p_b; \w_a,\w_b)
= \frac{1}{4N_c^2}\, \delta^{c'c}\,\delta^{d'd}\, \Bigl(\frac{\nslash_a}{2}\Bigr)_{\alpha\alpha'}
\, \Bigl(\frac{\nslash_b}{2}\Bigr)_{\beta\beta'}
\,f_{q/P}(p_a;\w_a)\,f_{q'/P}(p_b;\w_b)
\,.\end{equation}
%%%
In most cases of experimental interest, the observable is independent of $\w_{a,b}$ and the plus- and transverse components of the collinear momentum, which means we can integrate over these to obtain the standard parton distribution function~\cite{Collins:1981uw, Bauer:2002nz, Manohar:2003vb}
%%%
\begin{equation} \label{eq:xdep_qpdf}
f_{q/P}(x_a) =
\int\!\frac{\df^4 p_a}{(2\pi)^4}\,\Df\w_a\, f_{q/P}(p_a;\w_a)\,\delta(p_a^- - x_a\Ecm)
\,,\end{equation}
%%%
while everywhere else up to power corrections we can use
%%%
\begin{equation} \label{eq:pab_exp}
p_a = x_a\Ecm\,\frac{n_a}{2}
\,,\qquad
p_b = x_b\Ecm\,\frac{n_b}{2}
\,,\end{equation}
%%%
with $n_a = (1,\vn_{P_a})$ and $n_b = (1,\vn_{P_b})$ now aligned along the direction of the incoming protons.

While the dependence on $\w$ in our generalized distributions $f_{q/I}(p;\w)$ is not of relevance for most processes of interest, it describes the energy configuration of the remnant of the proton after the hard scattering. Thus, this matrix element provides a field-theoretical definition of the beam remnant. In particular, this means that the effect of the beam remnant is properly taken into account in our factorization proof in \sec{genericfactorization}.
In principle, operators $\cO_I$ with more than one collinear field in the directions $n_a$ and $n_b$ can be included as well, and would describe multiple scatterings of partons originating from the initial protons. These additional hard scatterings give rise to what is usually referred to as the underlying event~\cite{Sjostrand:1987su, Sjostrand:2004pf}. Thus, these effects are also taken into account in our factorization proof. Moreover, our formalism provides a field-theoretic basis to study the underlying event. The details are left for future work.

Since we only work to tree level in the matching from QCD to SCET, there is only a single soft function required for the process $q q' \to q q'$. After contracting with the color structures of the initial state function and the $q$ and $q'$ quark jet functions $J^{e'e}_{\gamma'\gamma}(p_1; \w_1)$ and $J^{f'f}_{\delta'\delta}(p_2; \w_2)$ (defined in the first line of \eq{Jquark_nonsinglet}), we obtain
%%%
\begin{align}
\label{eq:sqqqq}
S_{n_a n_b n_1 n_2}(\w_s)
&= \frac{2}{N_cC_F}\, \int\!\frac{\df^4 k_s'}{(2\pi)^4}\,\frac{\df^4 k_s}{(2\pi)^4}\, \Mae{0}{\cO_S^{\dagger ce\,df}(k_s')\, \deltai{s}\, \cO^{ec\,fd}_S(k_s)}{0}
\nn\\
&= \frac{2}{N_cC_F}\,
\Mae{0}{
\overline{T} \bigl[(\Yin^\dagger_{n_a} T^B Y_{n_1})^{ce}(0)\,(\Yin_{n_b}^\dagger T^B Y_{n_2})^{df}(0) \bigr]
\, \deltai{s}
\nn\\& \qquad\qquad\quad \times
T\bigl[(Y_{n_1}^\dagger T^A\Yin_{n_a})^{ec}(0)\,(Y_{n_2}^\dagger T^A\Yin_{n_b})^{fd}(0) \bigr]
}{0}
\,.\end{align}
%%%

%===============================================================================
\subsection{Generic Expression}
%===============================================================================

Combining Eqs.~\eqref{eq:dsigma_dw}, \eqref{eq:Jquark_nonsinglet}, \eqref{eq:C4}, \eqref{eq:generalqpdf} and \eqref{eq:sqqqq}, the $qq' \to qq'$ contribution to $\delta\sigma/\delta\w$ for 2-jet production can be written as
%%%
\begin{align}
\frac{\delta \sigma}{\delta \w}
&= \frac{1}{2 \Ecm ^2} \biggl(\prod_{i = a, b, 1,2}\int\!\frac{\df^4 p_i}{(2 \pi)^4}\, \Df \w_i \biggr)
\frac{1}{4 N_c^2}\, f_{q/P}(p_a; \w_a)\, f_{q'/P}(p_b; \w_b)\, J(p_1; \w_1)\, J(p_2; \w_2)\,
\nn\\ &\quad\times
\, \Hqq(p_a, p_b; p_1, p_2)
\,\frac{N_c C_F}{2}\int\!\Df \w_s\, S_{n_a n_b n_1 n_2}(\w_s) \,
\nn\\ &\quad\times
(2 \pi)^4 \delta^4 (p_a + p_b - P[\w_s] - p_1 - p_2)\,
\delta[\w - \w_a - \w_b - \w_1 - \w_2 - \w_s]
\,,\end{align}
%%%
where (at tree level in the matching)
%%%
\begin{align}
\Hqq(p_a, p_b; p_1, p_2)
&= \frac{g_s^4}{\hat t^2}\,\frac{1}{4}\,\tr[\nslash_a \gamma_\mu \nslash_1 \gamma_\nu]\,
   \frac{1}{4}\, \tr[\nslash_b \gamma^\mu \nslash_2 \gamma^\nu]
= \frac{2g_s^4}{\hat t^2}\, ( n_a \mcdot n_b\,\, n_1\mcdot n_2 + n_a\mcdot n_2 \,\, n_b \mcdot n_1)
\,.\end{align}
%%%

As discussed before, most observables are independent of the energy configurations $\w_a$ and $\w_b$, \ie, $f_O[\w_a + \w_b + \w_1 + \w_2 + \w_s] = f_O[\w_1 + \w_2 + \w_s]$. Therefore we can drop these beam remnant configuration in the $\delta$ functional for $\w$ and integrate over them in the parton distribution functions. Furthermore, inserting
%%%
\begin{equation}
1 = \Ecm^2 \int_0^1\!\df x_a \int_0^1\! \df x_b \, \delta(p_a^- - x_a \Ecm)\,\delta(p_b^- - x_b \Ecm)
\,,\end{equation}
%%%
and using \eqs{xdep_qpdf}{pab_exp}, the expression for $\delta\sigma/\delta\w$ becomes
%%%
\begin{align}
\frac{\delta \sigma}{\delta \w}
&= \frac{C_F}{16 N_c}
\int_0^1\!\df x_a\, \df x_b\,f_{q/P}(x_a)\, f_{q'/P}(x_b)
\int\!\frac{\df^4 p_1}{(2\pi)^4}\,\Df \w_1\, J(p_1; \w_1)
\int\!\frac{\df^4 p_2}{(2\pi)^4}\,\Df \w_2\, J(p_2; \w_2)
\nn\\&\quad\times
\, \Hqq\Bigl(\Ecm x_a\, \frac{n_a}{2}, \Ecm x_b\, \frac{n_b}{2}; p_1, p_2\Bigr)
\int\!\Df \w_s\, S_{n_a n_b n_1 n_2}(\w_s)
\nn\\&\quad\times
(2 \pi)^4 \delta^4\Bigl(\Ecm x_a\, \frac{n_a}{2} + \Ecm x_b\, \frac{n_b}{2} - P[\w_s] - p_1 - p_2\Bigr)
\,\delta[\w - \w_s - \w_1 - \w_2]
\,, \end{align}
%%%
where $n_{a,b} = (1, \vn_{P_{a,b}})$ are now aligned with the directions of the incoming protons. As in \sec{ee}, this can be simplified further by expanding the kinematics. After some algebra, we obtain
%%%
\begin{align} \label{eq:dsigma_dw_pp}
\frac{\delta \sigma}{\delta \w}
&= \int \frac{\df\Omega_p}{2 \pi}\int_0^1\!\df x_a\, \df x_b
\int\!\frac{\df p_1^+}{2 \pi}\, \Df\w_1\, J(p_1^+,p_1^-,{\bf n}_1; \w_1)
\int\!\frac{\df p_2^+}{2 \pi}\, \Df\w_2\, J(p_2^+, p_2^-,{\bf n}_2;\w_2)
\nn\\ & \times
f_{q/P}(x_a)\, f_{q'/P}(x_b)\,\frac{\df \sigma_0}{\df \cos \theta_p}
\int\!\Df\w_s\, S_{n_a n_b n_1 n_2}(\w_s)\, \delta[\w - \w_s - \w_1 - \w_2]
\,, \end{align}
%%%
where  the angular integral is defined in the center-of-mass frame of the partonic collision. Both the large $p_i^-$ components and the directions $\vn_i$ are functions of the partonic center-of-mass angular variables, $\Omega_p$, and the energy fractions of the incoming partons, $x_{a,b}$. They are defined by
%%%
\begin{align}
\label{eq:Eidef}
p_1^-(\Omega_p,x_a,x_b)  &= \frac{\Ecm}{2}\left[x_a(1+\cos \theta_p) + x_b(1-\cos \theta_p) \right]
\,,\nn\\
p_2^-(\Omega_p,x_a,x_b)  &= \frac{\Ecm}{2}\left[x_a(1-\cos \theta_p) + x_b(1+\cos \theta_p) \right]
\,,\nn\\
\vn_{i}(\Omega_p,x_a,x_b) &= \vn(\W_i)
\,,\end{align}
%%%
where $\W_i = (\theta_i, \phi_i)$ are given by
%%%
\begin{align}
\label{eq:nidef}
\cos\theta_1 &=  \frac{x_a(1+\cos \theta_p) - x_b(1-\cos \theta_p)}{x_a(1+ \cos \theta_p) + x_b(1-\cos \theta_p)}\,, &\phi_1 &= \phi_p
\nn\\
\cos\theta_2 &=  \frac{x_a(1-\cos \theta_p) - x_b(1+ \cos \theta_p)}{x_a(1-\cos \theta_p) + x_b(1+ \cos \theta_p)}\,, &\phi_2 &= \phi_p+\pi\,.
\end{align}
%%%
Finally, the differential cross section $\df \sigma_0 / \df \cos \theta_p $ is given by
%%%
\begin{equation}
\frac{\df \sigma_0}{\df \cos \theta_p} = \frac{\pi \alpha_s^2 C_F}{2 N_c}\, \frac{1}{x_a\, x_b\, \Ecm^2}\,
\frac{4+(1+ \cos \theta_p)^2}{(1-\cos \theta_p)^2}
\,,\end{equation}
%%%
which agrees with the well known expression in terms of the Mandelstam variables $\hat s,\hat t,\hat u$
%%%
\begin{equation}
\frac{\df \sigma_0}{\df \hat t}
= \frac{\pi \alpha_s^2 C_F}{N_c}\, \frac{\hat s^2 + \hat u^2}{\hat s^2\, \hat t^2}\,.
\end{equation}
%%%

%===============================================================================
\subsection{Jet Observables}
%===============================================================================

As an example how to use \eq{dsigma_dw_pp}, we derive a factorized cross section for infrared-safe cone jet observables, which was also studied in Ref.~\cite{Kidonakis:1998bk}. The required steps are very similar to the derivation given in \subsec{TwoJetCone}, and we only highlight the differences that arise from having protons in the initial state. First, $\delta\sigma / \delta \w$ depends on the parton distribution functions $f_{q/P}(x_{a})$ and $f_{q'/P}(x_{b})$. Second, while for $e^+ e^-$ collisions one often uses the variables $\theta$ and $\phi$ to denote the direction of jets, in $pp$ collisions it is more appropriate to use the rapidity instead of the angle $\theta$, due to the easier transformation properties under boosts along the beam direction. This gives cone jet functions $J_\cone$ that have exactly the same form as in \eq{dsigma_dw_pp} but use the corresponding cone projections in place of \eq{cone_J} to define the functionals $P_\cone[\w;\vj_i]$ in \eq{Piw_cone}. The final difference is that the cone soft function now explicitly depends on the orientation of the directions $\vn_i$ relative to the beam axis, since it contains Wilson lines in both the directions of the incoming protons and the outgoing jets. In particular, this implies that the nonperturbative physics described by this cone soft function depends on the rapidities of the outgoing jets. The experimental determination of the soft function is thus considerably more difficult for hadronic collisions than for $e^+ e^-$ collisions.

Keeping in mind these differences, we can follow the same steps as in \subsec{TwoJetCone} to obtain the factorization formula for generic two-jet observable using cone jets:
%%%
\begin{align}
\frac{\df \sigma}{\df O}
&= \int\!\frac{\df\Omega_p}{2\pi}\,\int_0^1 \! \df x_a\, \df x_b\,f_{q/P}(x_a)\, f_{q'/P}(x_b)\, \frac{\df\sigma_0}{\df \cos \theta_p} \nn\\
& \quad \times \int\!\df q_1^+\, \df q_1^- \, J_\cone(q_1^+,q_1^-)\int \df q_2^+\, \df q_2^- \, J_\cone(q_2^+,q_2^-)
\nn\\ &\quad\times
\int\!\df\ell_1^+\,\df\ell_2^+\,S^\cone_{\vn_1\vn_2}(\ell_1^+, \ell_2^+)\,
\delta\bigl(O - g_O(q_1^+ + \ell_1^+,q_1^-, q_2^+ + \ell_2^+,q_2^-,{\bf n}_1,{\bf n}_2)\bigr)
\,,\end{align}
%%%
where the cone soft function is now defined as
%%%
\begin{align}
S^\cone_{\vn_1\vn_2}(\ell_1^+, \ell_2^+) = \int\!\Df \w_s\,  S_{n_a n_b n_1 n_2}(\w_s)  \, \delta(\ell_1^+-n_1\mcdot P_\cone[\w_s;\vn_1])\, \delta(\ell_2^+-n_2\mcdot P_\cone[\w_s;\vn_2])
\,.\end{align}
%%%

As for $e^+ e^-$, many jet observables only depend on the large momentum components and the direction of the jets. In this case, we can perform the integrals over $p_i^+$ and $\ell_i^+$. Integrating over $\ell_i^+$ we define
%%%
\begin{equation}
\int \! \df \ell_1^+ \, \df \ell_2^+ \, S^\cone_{\vn_1,\vn_2}(\ell_1^+, \ell_2^+) \equiv S^\cone_{\vn_1,\vn_2}
\,,\end{equation}
%%%
which is now perturbatively calculable up to small power corrections. We obtain
%%%
\begin{align}
\frac{\df \sigma}{\df O}
&=
\int\!\frac{\df\Omega_p}{2\pi}\,\int_0^1 \! \df x_a\, \df x_b\,f_{q/P}(x_a)\, f_{q'/P}(x_b)\, \frac{\df\sigma_0}{\df \cos \theta_p}
\int\!\df q_1^- \, J(q_1^-)\int\!\df q_2^- \, J(q_2^-)
\nn\\ &\quad\times
S^\cone_{\vn_1,\vn_2}\, \delta\bigl(O - g_O(q_1^-, q_2^-,{\bf n}_1,{\bf n}_2)\bigr)
\,,\end{align}
%%%
where the jet functions integrated over $p_i^+$ are defined as in \eq{Jcone_integrated}.

%%%%%%%%%%%%%%%%%%%%%%%%%%%%%%%%%%%%%%%%%%%%%%%%%%%%%%%%%%%%%%%%%%%%%%%%%%%%%%%%
\section{Conclusions and Outlook}
\label{sec:conclusions}
%%%%%%%%%%%%%%%%%%%%%%%%%%%%%%%%%%%%%%%%%%%%%%%%%%%%%%%%%%%%%%%%%%%%%%%%%%%%%%%%

We have developed a new formalism for obtaining factorization theorems for almost any observable of interest at high energy colliders.
We argued that any observable differential cross section can be written in terms of two building blocks, a fully differential cross section describing the energy and momentum distribution of a given event, together with the restriction of how to obtain the desired observable from this distribution.
For events containing only massless particles in the final state, the only information required to define observables $O$ is the energy configuration $\w$ of the event, and we therefore focused on the cross section fully differential in $\w$, which we denoted as $\delta \sigma/\delta\omega$. By integrating this energy distribution with an appropriate functional $f_O[\w]$, the differential cross section $\df\sigma/\df O$ in any observable $O$ can be obtained.

Our main result is the proof of factorization for the fully differential cross section, $\delta\sigma/\delta\w$, using soft-collinear effective theory. It relies on the fact that $\delta \sigma/\delta\omega$ can be written directly in terms of a matrix element of well-defined operators in SCET using the energy flow operator. The linearity of the energy flow operator allowed us to factorize $\delta \sigma/\delta\omega$ into simpler building blocks, each of which is defined by matrix elements of operators in the effective theory and contains a single scale allowing for a systematic program of logarithmic resummation. After the factorized form of $\delta \sigma/\delta\omega$ for a given process is determined once and for all, it can be used to study the factorization properties of specific observables. The question of whether a given differential cross section $\df\sigma/\df O$ factorizes in the traditional sense depends on whether the form of $f_O[\w]$ is such that it smears the  individual matrix elements in $\delta \sigma/\delta\omega$ into objects that can be either calculated perturbatively or determined experimentally from other processes.

Using our formalism, we were able to directly study the factorization properties of the fully differential cross section, independent from the observable-specific functional $f_O[\w]$. While the question of whether the differential cross section in a given observable factorizes in the traditional sense still needs to be asked on an observable-by-observable basis, this disentanglement demonstrates to what length the steps taken in factorization proofs are observable independent. It turns out that it is the observable independent analysis that requires most of the calculational work. The fact that we can study factorization on an observable independent level could potentially be relevant for Monte Carlo event generation. It should be possible to make a connection between our factorized result for $\delta\sigma/\delta \w$ for generic $N$-jet production and the $N$-body partonic calculations that were introduced in Refs.~\cite{Bauer:2008qh, Bauer:2008qj} as input for an event generation framework. If so, our results could be used to provide improved theoretical inputs for event generation. However, more work in this direction is needed.

To demonstrate the simplicity with which factorization formulas for specific observables can be obtained from the factorized result for $\delta \sigma/\delta\omega$, we have applied our results to several simple observables in $e^+ e^- \to 2$ jets. We first reproduced the known results for event shape  and hemisphere mass distributions, and then obtained factorization formulas for generic observables defined in terms of the total jet momenta obtained from cone jet algorithms, which so far have not been studied in SCET.
We have also explored some of the issues arising in jet production in hadronic collisions by studying the partonic subprocess $qq' \to qq'$ using tree level matching from QCD onto SCET. In particular, we showed that the more complicated structure requires a soft function that is more complicated from the case of $e^+ e^-$ scattering. We also showed how parton distribution functions arise in our formalism, and commented on how it could be used to study beam remnants and underlying events.

It should be clear from these examples how our generic N-jet formalism can be applied to the study of observables in more complicated processes, such as processes with heavy vector bosons and more than two jets in the final state, which are crucial for many measurements at the upcoming LHC. It is these more complicated processes where the power of our new formalism becomes increasingly pronounced. While the number and complexity of Dirac and color structures grows quickly for any exhaustive study of factorization with two or more final state jets, the application of our formalism is straightforward and in fact facilitates recycling the bulk of the work needed or already known in the literature for a particular observable, to be used for other observables of interest.

%%%%%%%%%%%%%%%%%%%%%%%%%%%%%%%%%%%%%%%%%%%%%%%%%%%%%%%%%%%%%%%%%%%%%%%%%%%%%%%%
\begin{acknowledgments}
We would like to thank Chris Lee for useful conversations.
This work was supported in part by the Director, Office of
Science, Office of High Energy Physics of the U.S.\ Department of Energy under
the Contract DE-AC02-05CH11231.
C.W.B. acknowledges support from an DOE OJI award and an LDRD grant from LBNL.
\end{acknowledgments}

%%%%%%%%%%%%%%%%%%%%%%%%%%%%%%%%%%%%%%%%%%%%%%%%%%%%%%%%%%%%%%%%%%%%%%%%%%%%%%%%

\bibliographystyle{physrev4}
\bibliography{bibliography}

\end{document}